\documentclass[a4paper,11pt]{article}

\usepackage{jheppub} 

\usepackage[T1]{fontenc} 
\usepackage{txfonts}
\usepackage{braket}
\usepackage{latexsym}
\usepackage{wrapfig}
\usepackage{tabularx}
\usepackage{comment}
\newcommand{\der}{\partial}
\newcommand{\no}{\nonumber}

\newcommand{\im}{\mathrm{Im}}
\newcommand{\mass}{m_{\phi}}
\newcommand{\Mpl}{M_{\rm pl}}

\newcommand{\lag}{\mathcal{L}}
\newcommand{\lth}{\Lambda_{\rm EFT}}
\newcommand{\disc}{{\rm Disc}_s}

\title{Gravitational positivity bounds}

\author[a]{Junsei Tokuda,}
\author[b]{Katsuki Aoki,}
\author[c]{and Shin'ichi Hirano}

\affiliation[a]{Department of Physics, Kobe University,
 Kobe 657-8501, Japan}
\affiliation[b]{Center for Gravitational Physics, Yukawa Institute for Theoretical Physics, Kyoto University, Kyoto 606-8502, Japan}
\affiliation[c]{Division of Particle and Astrophysical Science,
Graduate School of Science, Nagoya University, Aichi 464-8602, Japan}

\emailAdd{jtokuda@amethyst.kobe-u.ac.jp}
\emailAdd{katsuki.aoki@yukawa.kyoto-u.ac.jp}
\emailAdd{hirano.shinichi@a.mbox.nagoya-u.ac.jp}

\abstract{We study the validity of positivity bounds in the presence of a massless graviton, assuming the Regge behavior of the amplitude. Under this assumption, the problematic $t$-channel pole is canceled with the UV integral of the imaginary part of the amplitude in the dispersion relation, which gives rise to finite corrections to the positivity bounds. We find that low-energy effective field theories (EFT) with ``wrong'' sign are generically allowed. The allowed amount of the positivity violation is determined by the Regge behavior. This violation is suppressed by $\Mpl^{-2}\alpha'$ where $\alpha'$ is the scale of Reggeization. This implies that the positivity bounds can be applied only when the cutoff scale of EFT is much lower than the scale of Reggeization. We then obtain the positivity bounds on scalar-tensor EFT at one-loop level. Implications of our results on the degenerate higher-order scalar-tensor (DHOST) theory are also discussed.}

\begin{document}
{\baselineskip0pt
\rightline{\baselineskip16pt\rm\vbox to-20pt{
           \hbox{KOBE-COSMO-20-13, YITP-20-98}
\vss}}%
}

\maketitle
\flushbottom

\section{Introduction}
It is well known that not every consistent-looking low-energy effective field theories (EFT) could be embedded into standard ultraviolet (UV) completions respecting some desired properties such as unitarity and causality. For non-gravitational theories, there is a clean argument that certain combinations of the EFT coefficients of higher-derivative corrections must possess correct sign if one requires the existence of standard UV completions. A classic example of the model which has a higher-derivative term with ``correct'' sign is \cite{Adams:2006sv} 
\begin{align}
\mathcal{L}_{\rm EFT}=-\frac{1}{2}(\der\phi)^2+\frac{c}{\Lambda^4}(\der\phi)^4+({\rm higher}-{\rm order}\quad\!\!{\rm terms})\,,\label{eq:classic}
\end{align}
with $c>0$. This model with $c\leq0$ does not admit a standard UV completion.\footnote{Strictly speaking, $c=0$ is allowed when the theory is free or when renormalizable couplings such as $\phi^4$ vertex in 4 dimensions exist.} It is also possible to constrain the coefficients of higher-order terms. These constraints are called positivity bounds in the literature. Positivity bounds are formulated in terms of the scattering amplitude of light fields at low energies, and hence they are free from field-redefinition ambiguities. In case of the model \eqref{eq:classic}, the lowest-order positivity bound is given by $\der_s^2F(s,0)|_{s=0}>0$. Here, $F(s,t)$ denotes a $2$ to $2$ scattering amplitude of $\phi$, and $(s, t, u)$ are Mandelstam variables. One can easily check that this bound indeed leads to the condition $c>0$ when considering the model \eqref{eq:classic}.

One of the most interesting consequence of the positivity bounds is that they relate a parameter space of phenomenological EFT to several properties of UV completion, where the former can be constrained experimentally. For instance, positivity bounds are applied to the standard model EFT in \cite{Zhang:2018shp, Remmen:2019cyz,Zhang:2020jyn}. If the violation of positivity bounds is confirmed in future experiments, the UV completion of the standard model sector will break at least one of the properties which are usually assumed as standard properties at UV.

It is now natural to think about applying positivity bounds to gravitational EFT. If they are valid, we can utilize them in order to extract the information of UV completion of gravity from EFT data. The positivity bounds with gravity have another important implication, actually: some mild version of Weak Gravity Conjecture could be derived if the positivity bounds were applicable to gravitational EFT \cite{Hamada:2018dde}. This also motivates us to carefully think about the validity of positivity bounds in the presence of a massless graviton. Unfortunately, it is known that the presence of a massless graviton might spoil the validity of positivity bounds. As we will explain in the main text in detail, the most problematic and unavoidable point which makes the positivity argument unclear is the fact that the $t$-channel exchange of massless graviton grows as $s^2$ at high energies. This $s^2$ growth means that $\der_s^2F(s,t)$ is singular in the forward limit $t\to0$. There are several discussions on the validity of positivity bounds in the presence of gravity \cite{Hamada:2018dde, Bellazzini:2019xts}. It is tempted to remove the singular $t$-channel pole and suppose that the positivity bounds would hold even after the pole is removed. However, the rigorous derivation of positivity bounds is still absent.\footnote{It is claimed in \cite{Bellazzini:2019xts} that positivity arguments could be derived by considering some compactification procedures, but some subtleties are argued in \cite{Alberte:2020jsk}.} In \cite{Hamada:2018dde}, the scattering of photons is considered in the presence of gravity, and it is argued that the positivity bound is satisfied for the $2$ to $2$ scattering of photons provided that the higher spin states which UV complete gravity are subdominant in this scattering. However, the sign of the contribution from the higher-spin states was not analyzed in detail.

It seems that more precise knowledge on UV completion would be required to derive the gravitational positivity bounds. It has been suggested that the Regge behavior of the amplitude, which is realized in string theory examples, may be deeply related to the causality of weakly-coupled UV completion of gravity (see {\it e.g.}, \cite{Camanho:2014apa, DAppollonio:2015fly}).

Motivated by the above observation, we discuss the validity of gravitational positivity bounds under the assumption that the scattering amplitude of the matter (a massive scalar $\phi$ in this study) is Reggeized at high energies. The point is that the Regge behavior actually allows us to compute the cancellation of the graviton $t$-channel pole at the level of twice subtracted dispersion relation. We can thus evaluate a finite residual contribution after the cancellation and discuss the residual effect on the gravitational positivity bounds.

We find that the Regge behavior will generically admit the violation of positivity bounds, and identify the form of the {\it negative} term which could be the origin of the positivity violation. The modulus of negativity is found to be suppressed by $\Mpl^{-2}{\alpha'}$, where $\alpha'$ is the scale of Reggeization,\footnote{This point has been already pointed out in \cite{Hamada:2018dde}.} and hence we are able to apply positivity bounds to gravitational EFT if its cutoff scale is much lower than the scale of Reggeization. We then apply our \emph{approximate} positivity bounds to scalar-tensor EFT at one-loop level. We finally discuss implications of the bounds on a phenomenological model of the scalar-tensor theory called degenerate higher-order scalar-tensor (DHOST) theory~\cite{Langlois:2015cwa, Crisostomi:2016czh, Achour:2016rkg, BenAchour:2016fzp}.\footnote{Recently, positivity bounds on Horndeski theories discussed in \cite{Melville:2019wyy}, and put strong constraints on these models combined with cosmological observations. We compare their results with ours in sec.~\ref{sec:pheno}.}

This paper is organized as follows: In sec.~\ref{sec:review1}, we briefly review the derivation of the bounds for non-gravitational scalar field theories. In sec.~\ref{sec:graviton}, we discuss the validity of positivity bounds in the presence of a massless graviton assuming the Regge behavior. 
In sec.~\ref{application}, we apply the positivity bounds on scalar-tensor EFT and discuss the implications on the DHOST theory in sec.~\ref{sec:pheno}. Readers who are interested in the application to the DHOST theory can check sec.~\ref{sec:pheno} to know the bounds on DHOST. Sec.~\ref{sec:concl} is devoted to the conclusion and discussions. Detail computations are presented in appendices.
Throughout this paper, we adopt the units with $c=\hbar=1$ and the notation $p^2\coloneqq\eta_{\mu\nu}p^\mu p^\nu=-\left(p^0\right)^2+\left(p^1\right)^2+\left(p^2\right)^2+\left(p^3\right)^2$. 

\section{Review: positivity bounds for scalar theories without gravity}\label{sec:review1}
In this section, we briefly review the derivation of positivity bounds for scalar EFT without gravity following \cite{Adams:2006sv, deRham:2017avq}. We assume that EFT contains only a single scalar field $\phi$ with mass $m_\phi$. Positivity bounds on this EFT are formulated in terms of the 2 to 2 scattering of a scalar field $\phi$.
As is well known, this scattering amplitude can be expressed as a function of Mandelstam invariants $(s,t,u)$ thanks to the Lorentz invariance. The definition of these variables are $s\coloneqq-(p_1+p_2)^2$, $t\coloneqq-(p_1-p_3)^2$, and $u\coloneqq-(p_1-p_4)^2$, where $(p_1, p_2)$ and $(p_3,p_4)$ denote a set of ingoing and outgoing four momenta, respectively. $s$ corresponds to the energy square in the center of mass frame. $s+t+u=4m_\phi^2$ is satisfied by definition, and 
hence the 2 to 2 scattering of a scalar field $\phi$ can be expressed as a function of $s$ and $t$. So, we refer to this scattering amplitude as $F(s,t)$ below. We may also write $u(s,t)\coloneqq 4m_\phi^2-s-t$.

\begin{figure}[tbp]
 \centering
  \includegraphics[width=.8\textwidth, trim=70 100 70 180,clip]{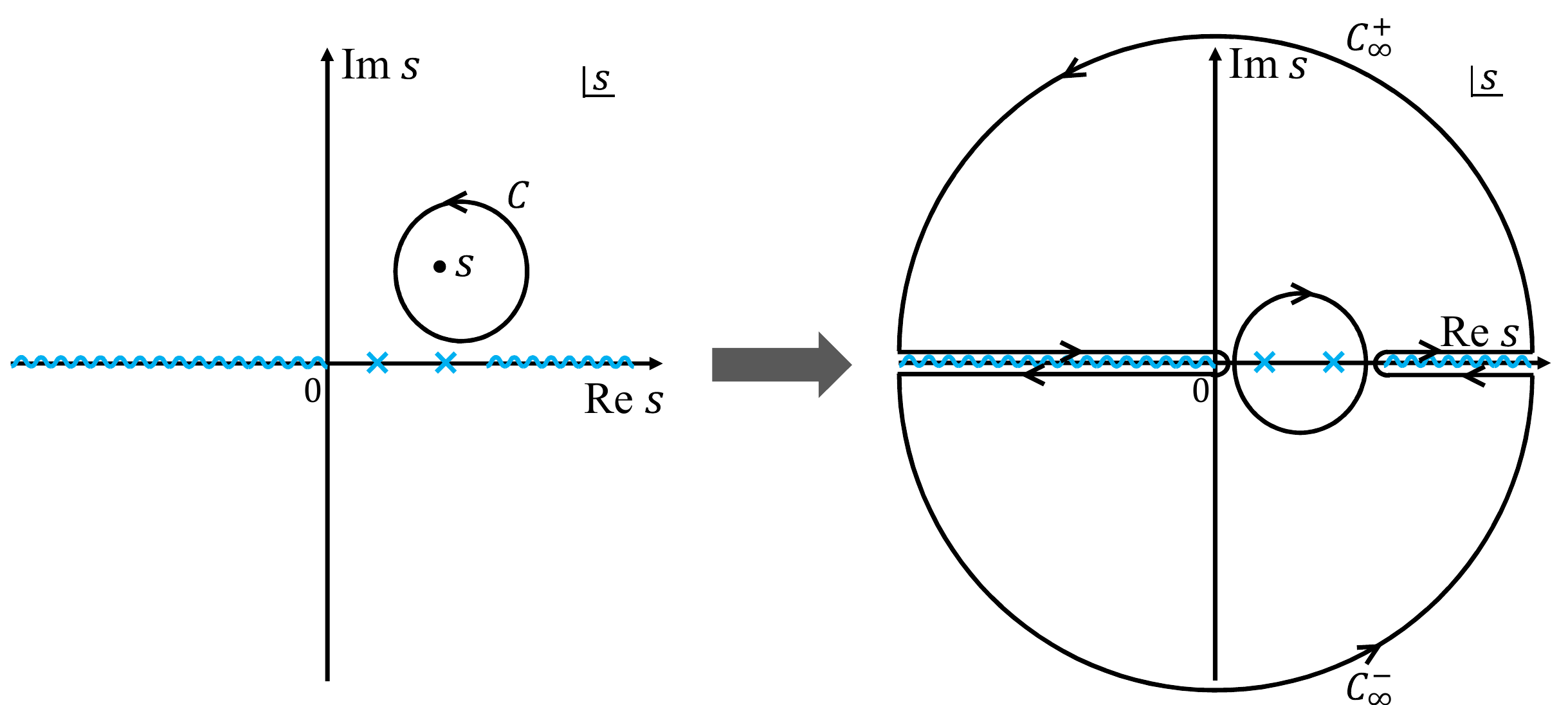}
 \caption{Analytic structure of $F(s,t)$ in the complex $s$-plane and the integration contours. The light-blue ``$\times$'' and the wavy lines are simple poles associated with a stable particle $\phi$ and branch cuts generated by loops, respectively. The integration contour $\mathcal{C}$ can be deformed into the one on the right panel, leading to the dispersion relation \eqref{eq:twdisp}.}
 \label{fig:deform} 
\end{figure}

To derive the positivity bounds, we start with expressing $F(s,t)$ by using the Cauchy's integral formula as
\begin{align}
F(s,t)=\left(s-2m_\phi^2+\frac{t}{2}\right)^{2}\oint_{\mathcal{C}}\frac{\mathrm d s'}{2\pi i}\frac{F(s',t)}{(s'-s)\left(s'-2m_\phi^2+\frac{t}{2}\right)^{2}}\,.
\end{align}
Here, $F(s,t)$ is holomorphic in $s$ inside the counterclockwise contour $\mathcal C$ as shown on the left panel of fig.~\ref{fig:deform}. Assuming that $F(s,t)$ is holomorphic in $s$ in the complex $s$-plane except poles and cuts, and $s\leftrightarrow u$ crossing symmetry, we can
deform the contour $\mathcal C$ as shown in fig.~\ref{fig:deform} to get the following twice subtracted dispersion relation for $t<4m_\phi^2$:
\begin{align}
F(s,t)=&\left[\frac{{\rm Res}_{s=m_\phi^2}F(s,t)}{s-m_\phi^2}+(s\leftrightarrow u(s,t))\right]+\sum_{k=0}^1a_k(t)s^k\no\\
&+\frac{2\left(\bar s+\frac{\bar t}{2}\right)^2}{\pi}\int^\infty_{4m_\phi^2}\mathrm{d}\mu\,\frac{\im\,F(\mu+i\epsilon,t)}{\left(\bar\mu+\frac{\bar t}{2}\right)\left[\left(\bar\mu+\frac{\bar t}{2}\right)^2-\left(\bar s+\frac{\bar t}{2}\right)^2\right]}\,,\label{eq:twdisp}
\end{align}
where the barred quantities are defined by $\bar Z\coloneqq Z-(4\mass^2/3)$, and ${\rm Res}_{s=m_\phi^2}\,F(s,t)$ denotes the residue of a pole at $s=m_\phi^2$. We did not write the explicit form of $a_k(t)$ because it is irrelevant in the analysis below. In eq.~\eqref{eq:twdisp},
we also used the following boundedness properties of $F(s,t)$ at large $|s|$
\begin{align}
\lim_{|s|\to\infty}\left|\frac{F(s+i\epsilon,t)}{s^2}\right|=0\quad{\rm for}\quad t<4m_\phi^2\,,\,t\neq m_\phi^2 \label{eq:nonfrois}
\end{align}
to ignore the contributions from infinity $C^\pm_{\infty}$.
This bound can be derived from the celebrated Froissart bound \cite{Froissart:1961ux, Martin:1962rt} in combination with the Phragm{\'{e}}n-Lindel{\"{o}}f theorem, and it reflects the locality of the theory.\footnote{The connection between the bound and the locality/non-locality of given theories is discussed in \cite{Tokuda:2019nqb}.} 

Let us define the pole subtracted amplitude $B(s,t)$ by  
\begin{align}
B(s,t)\coloneqq F(s,t)&-\left[\frac{{\rm Res}_{s=m_\phi^2}F(s,t)}{s-m_\phi^2}+\left(s\leftrightarrow u(s,t)\right)+\left(s\leftrightarrow t\right)\right]\,.\label{nongpolesubtdef1}
\end{align}
Then, defining $\tilde B(v,t)\coloneqq B(s,t)|_{s=v+2m_\phi^2-(t/2)}$ and $B^{(2N,0)}(t)\coloneqq\left.\der^{2N}_v\tilde B(v,t)\right|_{v=0}$, the dispersion relation \eqref{eq:twdisp} reads
\begin{align}
B^{(2,0)}(t)
=&\frac{4}{\pi}\int^\infty_{4m_\phi^2}\mathrm{d}\mu\,\frac{\im\, F\left(\mu+i\epsilon,t\right)}{\left(\mu-2m_\phi^2+\frac{t}{2}\right)^{3}}\,,
\end{align}
for $t<4m_\phi^2$, $t\neq m_\phi^2$. Generally, we can derive the $2N$-subtracted dispersion relation, leading to
\begin{align}
B^{(2N,0)}(t)
=&\frac{(2N)!2}{\pi}\int^\infty_{4m_\phi^2}\mathrm{d}\mu\,\frac{\im\, F\left(\mu+i\epsilon,t\right)}{\left(\mu-2m_\phi^2+\frac{t}{2}\right)^{2N+1}}
\,,\quad{\rm for}\quad N=1,2,\cdots\,.\label{subtdisp3}
\end{align}
According to the optical theorem, which is a consequence of unitarity, $\im\,F(s+i\epsilon,0)$ is strictly positive for $s\geq 4m_\phi^2$.\footnote{$\im\,F(s+i\epsilon,0)=0$ is allowed only when the theory is free \cite{deRham:2017imi}.} We thus obtain positivity bounds in the forward limit 
\begin{align}
B^{(2N,0)}(0)>0\,.\label{eq:bound1}
\end{align}
While the positivity bounds have been derived in the forward limit originally in \cite{Adams:2006sv}, later these bounds are extended to the $0<t<4m_\phi^2$ regime in \cite{deRham:2017avq}. 

The bound \eqref{eq:bound1} can be improved by computing known parts of $\im\,F(\mu+i\epsilon,0)$~\cite{Bellazzini:2016xrt, deRham:2017imi}.
Supposing that perturbative EFT computations are valid up to the scale $\lth$, we obtain the following improved positivity bounds for $t<4m_\phi^2$
\begin{align}
&B^{(2N,0)}_{\lth}(t)=\frac{(2N)!2}{\pi}\int^\infty_{\lth^2}\mathrm{d}\mu\,\frac{\im\, F\left(\mu+i\epsilon,t\right)}{\left(\mu-2m_\phi^2+\frac{t}{2}\right)^{2N+1}}>0\,,\label{eq:nogrposi}\\
&B^{(2N,0)}_{\lth}(t)\coloneqq B^{(2N,0)}(t)-\frac{(2N)!2}{\pi}\int^{\lth^2}_{4m_\phi^2}\mathrm{d}\mu\,\frac{\im\, F\left(\mu+i\epsilon,t\right)}{\left(\mu-2m_\phi^2+\frac{t}{2}\right)^{2N+1}}\,.\label{eq:defimpr}
\end{align}
From the definition of $\lth$ one can evaluate the left-hand side of eq.~\eqref{eq:nogrposi}. These bounds can be useful constraints on EFT.

It is also possible to derive bounds on the quantities $B^{(2N,M)}(t)\coloneqq (M!)^{-1}\der_v^{2N}\der_t^M\tilde B(v,t)|_{v=0}$. We briefly review the derivation of these bounds in the appendix~\ref{tderivext1}, and see \cite{deRham:2017avq} for details.

\section{Positivity bounds in the presence of gravity}\label{sec:graviton}
In the previous section \ref{sec:review1}, we reviewed the derivation of positivity bounds on scalar EFT. In this section, we discuss the effects of gravity to the positivity argument. For simplicity, we consider the system consisting of a massive scalar $\phi$ and a massless graviton $h$ which is defined as an excitation around the Minkowski spacetime, $g=\eta+h$ where $\eta$ is the Minkowski metric. Firstly, in sec.~\ref{subsec:problem}, we see how the problems arise once gravitational interactions are turned on. Next, in sec.~\ref{correcttreat}, we discuss the validity of positivity bounds assuming the Regge behavior of the amplitude. We find that unitarity will admit the violation of strict positivity. In sec.~\ref{subsec:stringeg}, we analyze the superstring amplitude as an explicit example which indeed violates the strict positivity. In sec.~\ref{appposi}, we put the upper bound on the modulus of {\it negativity} and discuss under which condition the positivity bounds are useful even in the presence of a massless graviton.

\subsection{Unavoidable problems caused by gravitons}\label{subsec:problem}

Once turning on the gravity, we immediately encounter several problems. Firstly, the presence of a massless mediator generally changes the analytic structure of the amplitude in the complex $s$-plane: branch cuts associated with the loops of massless particles prevent us from relating the upper-half plane to the lower-half plane. To avoid this issue, in this study, we assume that  
the graviton loop corrections to $\phi\phi\to\phi\phi$ scattering can be neglected.  
Below we refer to the $\phi\phi\to\phi\phi$ scattering amplitude with neglecting the graviton loops as $F(s,t)$. Secondly, the presence of a massless mediator invalidates the rigorous derivation of the constraints on the high energy behavior \eqref{eq:nonfrois} on which the derivation of positivity bounds crucially relies. This is also an obvious caveat for the validity of positivity bounds, however, in this study we assume that the similar boundedness property holds:
\begin{align}
\lim_{|s|\to\infty}\left|\frac{F(s+i\epsilon,t<0)}{s^2}\right|=0\,. \label{compfrois}
\end{align}
Here we omit $t=0$ at which the graviton $t$-channel pole is located and thus $F(s,t)$ is divergent. These two assumptions are just assumptions, but they are indeed satisfied in known examples, such as the superstring amplitude which we will analyze in sec.~\ref{subsec:stringeg}. 
Under these conditions, the derivation of the $2N$-subtracted dispersion relation for $t<0$ goes in parallel to the discussion in sec.~\ref{sec:review1}, leading to the following equations for $t<0:$
\begin{subequations}
\label{Tchanposi}
\begin{align}
&B_{\lth}^{(2,0)}(t)
=\frac{4}{\pi}\int^\infty_{\lth^2}\mathrm{d}\mu\,\frac{\im\, F\left(\mu+i\epsilon,t\right)}{\left(\mu-2m_\phi^2+\frac{t}{2}\right)^{3}}+\left[\der_v^{2}\left(\frac{\left.{\rm Res}_{t=0}F(s,t)\right|_{s=v+2m_\phi^2-(t/2)}}{-t}\right)\right]_{v=0}\,,\label{tchanposi1}\\
&B_{\lth}^{(2N,0)}(t)
=\frac{(2N)!2}{\pi}\int^\infty_{\lth^2}\mathrm{d}\mu\,\frac{\im\, F\left(\mu+i\epsilon,t\right)}{\left(\mu-2m_\phi^2+\frac{t}{2}\right)^{2N+1}}\quad({\rm for}\,\,N\geq 2)\,.\label{tchanposi1a}
\end{align} 
\end{subequations}
Here, the definition of $B^{(2N,0)}_{\lth}(t)$ is completely the same as the one in the previous section after replacing the definition of the pole subtracted amplitude \eqref{nongpolesubtdef1} by
\begin{align}
B(s,t)\coloneqq F(s,t)&-\left[\left(\frac{{\rm Res}_{s=m_\phi^2}F(s,t)}{s-m_\phi^2}+\frac{{\rm Res}_{s=0}F(s,t)}{s}\right)+\left(s\leftrightarrow u(s,t)\right)+\left(s\leftrightarrow t\right)\right]\,.\label{polesubtdef1}
\end{align}
The poles associated with both the scalar and the graviton are subtracted.
Let us compare eqs.~\eqref{Tchanposi} with \eqref{eq:nogrposi}. One can see that the dispersion relation with more than two subtractions \eqref{tchanposi1a} are formally the same as the one without gravity because ${\rm Res}_{t=0}F(s,t)\sim s^2$ so that the last term of \eqref{tchanposi1} disappears. Therefore, after taking the $t\to0$ limit starting from $t<0$ side and using the positivity of $\im\, F(s,-0)$, one can derive improved positivity bounds $B^{(2N,0)}_{\lth}(0)>0$ with $N\geq2$. On the other hand, the twice subtracted dispersion relation \eqref{tchanposi1} takes the different form from the one without gravity \eqref{eq:nogrposi}. The contribution of the graviton $t$-channel exchange remains on the r.h.s.~of \eqref{tchanposi1}. This makes the positivity of the r.h.s.~of \eqref{tchanposi1} subtle: in the limit $t\to-0$ the second term of the r.h.s.~of \eqref{tchanposi1} is divergent as $t^{-1}$, while the l.h.s.~is finite because it is defined in terms of the pole-subtracted amplitude $B(s,t)$. This means that the first term of the r.h.s.~of \eqref{tchanposi1} must be also divergent as $t^{-1}$ in the $t\to-0$ limit and then the r.h.s. becomes finite as a result of ``$\infty-\infty$''. We need to explicitly compute the cancellation of divergences in the $t\to-0$ limit and check the sign of finite residuals in order to investigate the validity of the lowest-order positivity $B^{(2,0)}_{\lth}(0)>0$ in the presence of gravity.

\subsection{Violation of strict positivity and the Regge behavior}\label{correcttreat}
The necessity of the $\infty-\infty$ cancellation of the r.h.s.~of \eqref{tchanposi1} in the forward limit has an interesting implication on the growth rate of $\im\,F(s+i\epsilon,t)$.
The imaginary part $\im\, F(s+i\epsilon,t)$ itself does not have any singular term in the limit $t\to-0$.\footnote{$\im\,F(s+i\epsilon,t)$ contains the term proportional to $\delta(t)$ which expresses the existence of the graviton $t$-channel exchange. Nonetheless, our dispersion relation does not have $\delta(t)$ because the dispersion relation is derived for a fixed {\it negative} $t$. The limit $t\to-0$ is taken after evaluating the cancellation of singular terms of the order of $t^{-1}$.} Hence, $\lim_{t\to-0}\im\, F(s+i\epsilon,t)$ must grow as fast as $s^2$ at high energies and then the UV integral of $\im\, F(s+i\epsilon,t)$ is divergent in the forward limit. This immediately suggests that the UV information of $\im\, F(s+i\epsilon,t)$ should be required to compute the $\infty-\infty$ cancellation. 

Then, how does the $s^2$ growth of $\im\, F(s,-0)$ correctly reproduce the graviton $t$-channel pole in \eqref{tchanposi1}? It is known that (and as we will see below) the reproduction of the graviton $t$-channel pole in the dispersion relation can be consistently implemented by the following Regge behavior at sufficiently high energies for negative, but in the vicinity of $t=0$
\begin{align}
\lim_{s\to\infty}\im\, F(s,t)=
f(t)\left(\frac{\alpha's}{4}\right)^{2+j(t)}
\,.\label{regge2}
\end{align}
Here $f(t)$ and $j(t)$ are regular functions in the vicinity of $t=0$. The $s^2$ growth of $\im\, F(s,-0)$ implies $j(0)=0$, while the bound \eqref{compfrois} requires $j(t)<0$ for negative $t$. The scale $\alpha'>0$, whose mass dimension is $-2$, is determined by the scale where the amplitude is Reggeized.  For instance, in string theory an infinite tower of massive higher spin states gives rise to the Regge behavior, and in this case the scale $\alpha'$ is nothing but the usual $\alpha'$ parameter in string theory: the amplitude is Reggeized above the string scale $M_{\rm s}=\alpha'^{-1/2}$.\footnote{Strictly speaking, in the tree-level string amplitude, the imaginary part of the amplitude consists of sequences of an infinite number of delta functions. Therefore, to relate it to the Regge behavior of the form \eqref{regge2}, it is necessary to perform some kind of ``smoothing'' of the amplitude: see~\ref{subsec:stringeg}.} We will study the string amplitude as an example in sec.~\ref{subsec:stringeg}. Since several works ({\it e.g.} \cite{Camanho:2014apa, DAppollonio:2015fly}) suggest that the Reggization of the amplitude will be deeply related to the causality in weakly-coupled UV completion of gravity, the Regge behavior \eqref{regge2} would be one of the most plausible assumption to compute the UV integral.

The Regge behavior eq.~\eqref{regge2} is just an asymptotic behavior of the amplitude in the high-energy limit. In eq.~\eqref{regge2}, we have dropped the terms whose growth rate is slower than $s^2$ in the Regge limit when $t=-0$. In the main text, we will assume the following simplest form 
\begin{align}
\im\, F(s,t)=f(t)\left(\frac{\alpha's}{4}\right)^{2+j(t)}\left[1+\mathcal{O}\left(\frac{1}{\alpha's}\right)\right]
\,,\label{regge3}
\end{align}
where sub-leading corrections appear in the form of $\mathcal{O}\,\bigl((\alpha's)^{-1}\bigr)$.\footnote{One can consider more generic form, although it will not change the main result of this paper: see app.~\ref{reggesample}. We note here that such a modification of sub-leading terms in \eqref{regge3} can give $\mathcal{O}(1)$ corrections to the finite pieces of eq.~\eqref{subtdisp8} which we will derive soon, but the sign of such corrections is not fixed by unitarity.}  
Because this scale is related to UV physics, we assume that $\lth^2\ll {\alpha'}^{-1}$ below, while in principle ${\alpha'}^{-1}$ can be comparable to the EFT cutoff scale $\lth^2$.

From now let us demonstrate that the Regge behavior \eqref{regge3} correctly reproduces the graviton $t$-channel pole and evaluate the finite terms of the r.h.s.~of \eqref{tchanposi1}. Using \eqref{regge3}, the first term of the r.h.s.~of eq.~\eqref{tchanposi1} can be evaluated as
\begin{align}
\int^\infty_{\lth^2}\mathrm{d}\mu\,\frac{\im\, F\left(\mu,t\right)}{\left(\mu-2m_\phi^2+\frac{t}{2}\right)^{3}}=\int^{M^2_*}_{\lth^2}\mathrm{d}\mu\,\frac{\im\, F\left(\mu,t\right)}{\left(\mu-2m_\phi^2+\frac{t}{2}\right)^{3}}+D\left(t;M_*^2\right)\,,\label{cutint1}
\end{align}
where
\begin{align}
D\left(t;M_*^2\right)\coloneqq\int^\infty_{M^2_*}\mathrm{d}\mu\,\frac{\im\, F\left(\mu,t\right)}{\left(\mu-2m_\phi^2+\frac{t}{2}\right)^{3}}\simeq\int^\infty_{M^2_*}\mathrm{d}\mu\,\frac{\im\, F\left(\mu,t\right)}{\left(\mu+\frac{t}{2}\right)^{3}}\,.\label{eq:defdiv}
\end{align}
We assumed that the imaginary part of the amplitude is well approximated by the Regge behavior \eqref{regge3} above a mass scale $M_*$. $D\left(t;M_*^2\right)$ is divergent in the $t\to-0$ limit, while it is finite for  $t<0$. This divergent piece must cancel the $t^{-1}$ singularity of the second term. This can be seen explicitly as follows: substituting \eqref{regge3} into \eqref{eq:defdiv}, $D\left(t;M_*^2\right)$ can be evaluated as
\begin{align}
D\left(t;M_*^2\right)=\frac{-f(t)\alpha'^2}{16}\left[\frac{1}{j(t)}\left(\frac{\alpha'M_*^2}{4}\right)^{j(t)}+\mathcal{O}\left(\frac{1}{\alpha'M_*^2}\right)\right]+\mathcal{O}(t)\,.\label{uvint1}
\end{align}
Here, $\mathcal{O}\left((\alpha'M_*^2)^{-1}\right)$ term comes from the sub-leading terms in \eqref{regge3}, which is finite in the forward limit. Taking $j(0)=0$ into account, eq.~\eqref{uvint1} reduces to
\begin{align}
D\left(t;M_*^2\right)
=\frac{-f\alpha'^2}{16}\left[\frac{1}{tj'}+\frac{f'}{fj'}+\ln\left(\frac{\alpha'M^2_*}{4}\right)-\frac{j''}{2(j')^2}+\mathcal{O}\left(\frac{1}{\alpha'M_*^2}\right)\right]+\mathcal{O}(t) \label{uvint2}
\end{align}
for a negative but in the vicinity of $t=0$. Here, we refer to $f(0)$ as $f$, and the derivatives of $f(t)$ or $j(t)$ with respect to $t$ evaluated at $t=0$ as $f'$, $j'$, or $j''$, respectively for notational simplicity. It is now clear that $D(t;M_*^2)$ is divergent as $t^{-1}$.\footnote{We also used $\der_t j|_{t=0}\neq0$ to derive eq.~\eqref{uvint2}, strictly speaking. However, the divergent behavior of $D(t;M_*^2)$ in the forward limit will be different from $t^{-1}$ when $\der_t j|_{t=0}=0$, which is inconsistent with the finiteness of the r.h.s.~of \eqref{tchanposi1}.} Finiteness of the r.h.s.~of eq.~\eqref{tchanposi1} requires\footnote{Reproduction of $t$-channel pole at the level of dispersion relation by the Regge behavior is a known phenomenon. See for instance the appendix A.2 of \cite{Caron-Huot:2016icg}.}
\begin{align}
\frac{f\alpha'^2}{4\pi j'}=-\left[\der_v^{2}\left(\left.{\rm Res}_{t=0}F(s,t)\right|_{s=v+2m_\phi^2-(t/2)}\right)\right]_{v=0}\,,\label{tpolecancel1}
\end{align}
which determines the value of $f\alpha'^2/j'$ once the EFT Lagrangian is fixed. In our case, the coupling strength of $\phi\phi h$ is $M_{\rm pl}^{-1}$, and hence eq.~\eqref{tpolecancel1} gives rise to
\begin{align}
\frac{f\alpha'^2}{4\pi j'}\sim  M^{-2}_{\rm pl}\,.\label{tpolecancel2}
\end{align}
Using eqs.~\eqref{cutint1}, \eqref{uvint2}, and \eqref{tpolecancel1}, eq.~\eqref{tchanposi1} reads
\begin{align}
&B_{\lth}^{(2,0)}\left(0\right)=
\frac{4}{\pi}\int^{M_*^2}_{\lth^2}\mathrm{d}\mu\,\frac{\im\, F\left(\mu,-0\right)}{\left(\mu-2m_\phi^2\right)^{3}}-\frac{f\alpha'^2}{4\pi}\left[\frac{f'}{fj'}+\ln\left(\frac{\alpha'M^2_*}{4}\right)\right]+\frac{f\alpha'^2}{4\pi}\left[\frac{j''}{2(j')^2}+\mathcal{O}\left(\frac{1}{\alpha'M_*^2}\right)\right]\,,\label{subtdisp8}
\end{align}
after taking the limit $t\to-0$. Note that one can show that the r.h.s.~of \eqref{subtdisp8} is $M_*$-independent up to $\mathcal{O}\left((\alpha'M_*^2)^{-1}\right)$ terms, by using eq.~\eqref{regge3}. The first term on the r.h.s.~of \eqref{subtdisp8} is ensured to be positive as a consequence of unitarity as usual. In addition to this strictly positive term, we find the finite residuals of $t$-channel pole cancellation. Specifically, {\it unitarity concludes that the second term must be negative}:
\begin{align}
-\frac{f\alpha'^2}{4\pi}\left[\frac{f'}{fj'}+\ln\left(\frac{\alpha'M^2_*}{4}\right)\right]\simeq-\frac{4}{\pi M_*^4j'}\left.\der_t\,\im\,F\left(M_*^2,t\right)\right|_{t=0}<0\,.\label{eq:negativity}
\end{align}
This implies that the finite residuals of $t$-channel pole cancellation on the r.h.s.~of \eqref{subtdisp8} can be negative in general, and must be negative when the leading Regge trajectory is linear {\it i.e.} $j''=0$. This is indeed the case for known examples as we will discuss in sec.~\ref{subsec:stringeg}.
Our analysis suggests that usual properties of $S$-matrix such as unitarity, analyticity, and the boundedness property at high energies \eqref{compfrois} alone will be insufficient to ensure the strict positivity $B^{(2,0)}_{\lth}(0)>0$: rather unitarity admits $B^{(2,0)}_{\lth}(0)\leq0$. We will discuss the known amplitude satisfying $B^{(2,0)}_{\lth}(0)=0$ in the next sec.~\ref{subsec:stringeg}.

\subsection{An example: superstring amplitude}\label{subsec:stringeg}
In the previous section~\ref{correcttreat}, we discussed the importance of the Regge behavior of the amplitude to cancel the graviton $t$-channel pole. We then observed that the lowest-order positivity bound is {\it not} necessarily satisfied. As a specific example of the amplitude exhibiting the Regge behavior which is consistent with unitarity and analyticity but violates strict positivity, let us consider for a while the $2$ to $2$ amplitude for NS-NS bosons of type-II superstring theory of the following form:
\begin{align}
F(s,t)=\left.-P(s,t)\frac{\Gamma\left(-\frac{\alpha's}{4}\right)\Gamma\left(-\frac{\alpha't}{4}\right)\Gamma\left(-\frac{\alpha'u}{4}\right)}{\Gamma\left(1+\frac{\alpha's}{4}\right)\Gamma\left(1+\frac{\alpha't}{4}\right)\Gamma\left(1+\frac{\alpha'u}{4}\right)}\right|_{u=u(s,t)}\,,\label{eq:example0}
\end{align}
with $s+t+u(s,t)=0$. Here, $P(s,t)$ denotes the factor depending on the external polarizations. In our case, 
 $P(s,t)=A(s^2u^2+t^2u^2+s^2t^2)|_{u=u(s,t)}$ with a proportionality constant $A>0$. An explicit expression of $A$ is irrelevant in the analysis below. This amplitude contains an infinite number of simple poles. For instance, there are $t$-channel poles at $t=0, 4J/\alpha'$ with $J=1,2,\cdots$. A pole at $t=0$ expresses the contribution of $t$-channel graviton exchange: the corresponding residue is 
\begin{align}
{\rm Res}_{t=0}\,F(s,t)=-\frac{64As^2}{\pi\alpha'^3}\,.\label{eq:string_res}
\end{align}
 Other poles at $t=4J/\alpha'$ with $J=1,2,\cdots$ are massive higher-spin states, which are called Regge states. The residue of a pole at $t=4J/\alpha'$ grows as $s^{(J+2)}$ at high energies, indicating the presence of a spin-$(J+2)$ state. These states are responsible for the Regge behavior of the amplitude. 

To discuss the high-energy behavior, it may be useful to rewrite \eqref{eq:example0} by using an equality $\Gamma(1-x)\Gamma(x)\sin(\pi x)=\pi$ as
\begin{align}
F(s,t)=\frac{-P}{\left[\Gamma\left(1+\frac{\alpha't}{4}\right)\right]^2}\left[\cot\left(\frac{\pi\alpha't}{4}\right)+\cot\left(\frac{\pi\alpha's}{4}\right)\right]\left(\frac{\Gamma\left(\frac{\alpha's}{4}+\frac{\alpha't}{4}\right)}{\Gamma\left(1+\frac{\alpha's}{4}\right)}\right)^2\,.\label{eq:example1}
\end{align}
Computing the imaginary part of \eqref{eq:example1}, we have
\begin{align}
\left.\im\,F(s+i\epsilon,t)\right|_{s>0}=\sum_{J=1}^\infty\frac{4P}{\alpha'}\left(\frac{\Gamma\left(J+\frac{\alpha't}{4}\right)}{\Gamma\left(1+\frac{\alpha't}{4}\right)\Gamma(1+J)}\right)^2\delta\left(s-\frac{4J}{\alpha'}\right)\,,\label{eq:reggeresi1}
\end{align}
when $0<|t|\alpha'<s\alpha'$. An infinite number of $u$-channel poles appear for $s\leq0$. 
From this one can compute 
\begin{align}
\int^\infty_{\lth^2}\mathrm{d}\mu\,\frac{\im\,F\left(\mu,t\right)}{\left(\mu+\frac{t}{2}\right)^3}&=\frac{16A}{\alpha'^2}\sum^\infty_{J=1}\frac{1}{J}\left(\frac{\Gamma\left(J+\frac{\alpha't}{4}\right)}{\Gamma\left(1+\frac{\alpha't}{4}\right)\Gamma\left(J\right)}\right)^2+\mathcal{O}(t)
=-\frac{32A}{\alpha'^3t}+\mathcal{O}(t)\,.\label{eq:singular1}
\end{align}
Here, we regard the Regge states as heavy states which do not appear in EFT, and hence we choose $\lth^2<4/\alpha'$ where $4/\alpha'$ is the mass square of the lightest Regge states. This correctly reproduces the graviton $t$-channel pole: substituting this and \eqref{eq:string_res} into eq.~\eqref{tchanposi1}, we have
\begin{align}
B^{(2,0)}_{\lth}(0)
=0\,.\label{eq:zero1}
\end{align}
This exact cancellation between the graviton $t$-channel pole and the contribution from the Regge states is also obvious by directly expanding the pole subtracted amplitude in terms of $s$ at $s=0$ in the forward limit.\footnote{Note that the sign of low-energy expansion coefficients of tree-level superstring amplitudes has been investigated in \cite{Green:2019tpt}, for instance.} We emphasize here that each Regge state indeed gives the strictly positive contribution to $B^{(2,0)}_{\lth}(0)$ as expected from unitarity: for instance, the contribution from the Regge states with mass square $4J/\alpha'$ to $B^{(2,0)}_{\lth}(0)$ is $16A/(\alpha'^2J)$. However, the contributions from an infinite tower of Regge states to $B^{(2,0)}_{\lth}(0)$ are divergent, and are {\it exactly} canceled with the graviton $t$-channel pole.

\begin{figure}[tbp]
 \centering
  \includegraphics[width=.8\textwidth, trim=190 270 170 130,clip]{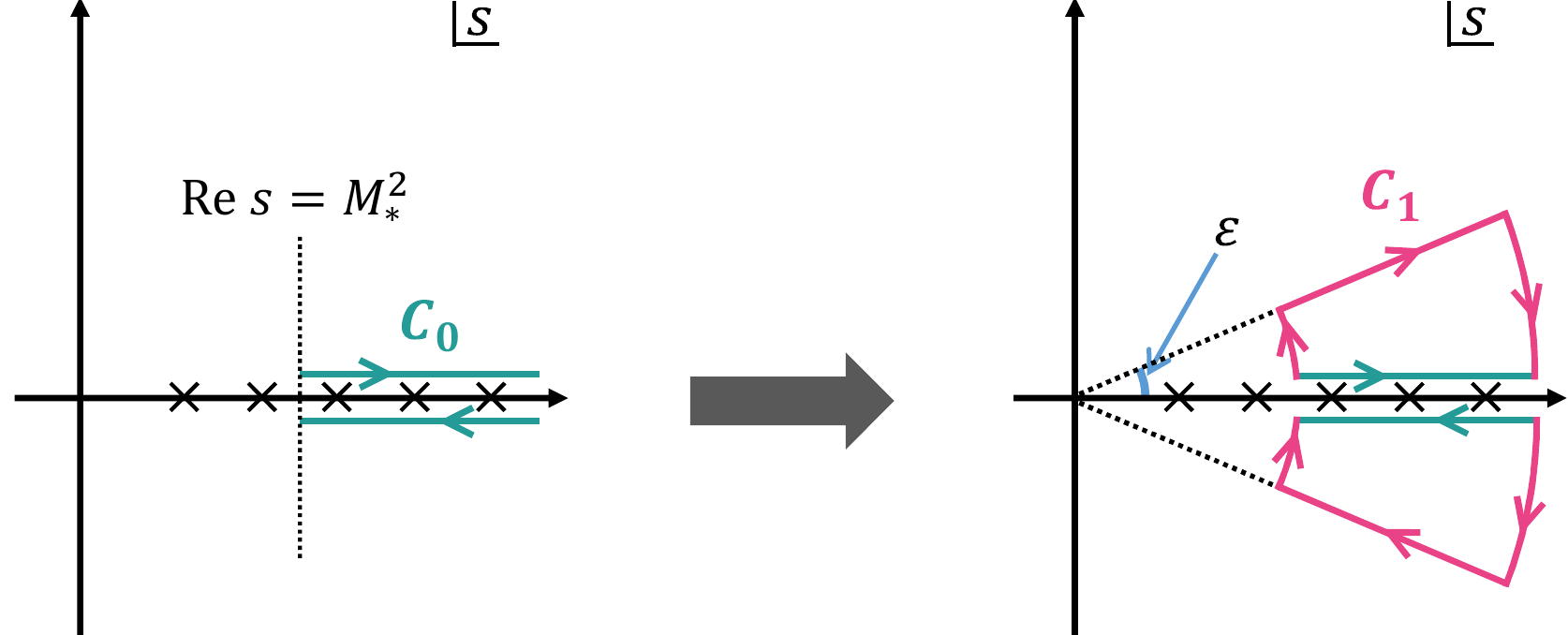}
 \caption{The deformation of the integration contours, parameterized by an angle $\varepsilon>0$. Simple poles associated with the Regge states are denoted by $\times$.}
 \label{smearing} 
\end{figure}

From now on, we discuss this cancellation of the graviton $t$-channel pole in the language of the previous section \ref{correcttreat}. For this purpose, 
we may need to consider the ``smoothed'' amplitude. One way to perform the ``smoothing'' is to consider $F(se^{i\theta},t)$ with $0<\theta\ll1$ instead of $F(s,t)$. Because this function is evaluated in the regions far away from the real $s$-axis for large ${\rm Re}\,s$, the contributions from an infinite number of poles to $F(se^{i\theta},t)$ are smoothed at high-energies.  From eq.~\eqref{eq:example1}, the smoothed amplitude is
\begin{align}
F(se^{i\theta},t)
&\approx\frac{-A\left(\alpha'/4\right)^{-2+\frac{\alpha'}{2}}}{\left[\Gamma\left(1+\frac{\alpha't}{4}\right)\right]^2}\left[\cot\left(\frac{\pi\alpha't}{4}\right)+\cot\left(\frac{\pi\alpha'se^{i\theta}}{4}\right)\right]\left(se^{i\theta}\right)^{2+\frac{\alpha't}{2}}\,.\label{eq:smooth1}
\end{align}
We used the Stirling's formula $\Gamma(z)\approx \sqrt{\frac{2\pi}{z}}z^ze^{-z}$ for ${\rm Re}\,z\gg 1$ and neglected the terms suppressed by $(\alpha'|s|)^{-1}$ or $|t/s|$.
Now, let us compute $D(t;M_*^2)$ which is defined by \eqref{eq:defdiv} by using \eqref{eq:smooth1}.  Firstly we rewrite \eqref{eq:defdiv} by deforming the integration contour as shown in fig.~\ref{smearing} to get 
\begin{align}
D\left(t;M_*^2\right)&\simeq \int^\infty_{M^2_*}\mathrm{d}\mu\,\frac{\im\,F(\mu,t)}{\mu^3}=\frac{1}{2i}\int_{\mathcal{C}_0}\mathrm{d}s\frac{F(s,t)}{s^3}=\frac{1}{2i}\int_{\mathcal{C}_1}\mathrm{d}s\frac{F(s,t)}{s^3}\no\\
&=\frac{1}{2i}\int^\infty_{M^2_*}\frac{\mathrm{d}\mu}{\mu}\,\disc\left[\frac{F(\mu e^{i\varepsilon},t)}{\mu^2e^{2i\varepsilon}}\right]+\frac{1}{2}\int^\varepsilon_0\mathrm{d}\theta\, \left[\frac{F\left(M_*^2e^{i\theta},t\right)}{M_*^4e^{2i\theta}}+\frac{F\left(M_*^2e^{-i\theta},t\right)}{M_*^4e^{-2i\theta}}\right]\,,\label{eq:defdiv2}
\end{align}
where $\disc$ denotes the discontinuity across the real $s$-axis: 
\begin{align}
\disc\left[\frac{F(s e^{i\varepsilon},t)}{s^2e^{2i\varepsilon}}\right]\coloneqq \frac{F(s e^{i\varepsilon},t)}{s^2e^{2i\varepsilon}}-\frac{F(s e^{-i\varepsilon},t)}{s^2e^{-2i\varepsilon}}\,.
\end{align}
From eq.~\eqref{eq:smooth1}, the discontinuity is given by
\begin{align}
&\disc\left[\frac{F(s e^{i\varepsilon},t)}{s^2e^{2i\varepsilon}}\right]\approx\no\\
& \frac{2iA\left(\alpha'/4\right)^{-2+\frac{\alpha't}{2}}s^{\frac{\alpha't}{2}}}{\left[\Gamma\left(1+\frac{\alpha't}{4}\right)\right]^2}\left[\cos\left(\frac{\varepsilon\alpha't}{2}\right)-I(s,\varepsilon)\cos\left(\frac{\varepsilon\alpha't}{2}\right)-\left(\cot\left(\frac{\pi\alpha't}{4}\right)+R(s,\varepsilon)\right)\sin\left(\frac{\varepsilon\alpha't}{2}\right)\right]\,.\label{eq:stringdisc1}
\end{align}
Here, $I(s,\varepsilon)$ and $R(s,\varepsilon)$ are defined by $\cot(\pi\alpha's e^{i\theta}/4)=R(s,\theta)+i(I(s,\theta)-1)$. Explicitly,
\begin{subequations}
\label{eq:IR}
\begin{align}
&I(s,\varepsilon)=\frac{\cos^2\left(\frac{\pi\alpha's}{4}\cos\varepsilon\right)+\cosh^2(\frac{\pi\alpha's}{4}\sin\varepsilon)-\cosh(\frac{\pi\alpha's}{4}\sin\varepsilon)\sinh(\frac{\pi\alpha's}{4}\sin\varepsilon)}{\cos^2(\frac{\pi\alpha's}{4}\cos\varepsilon)-\cosh^2(\frac{\pi\alpha's}{4}\sin\varepsilon)}\,,\\
&R(s,\varepsilon)=\frac{\frac{1}{2}\sin\left(\frac{\pi\alpha's}{2}\cos\varepsilon\right)}{\sin^2\left(\frac{\pi\alpha's}{4}\cos\varepsilon\right)+\sinh^2\left(\frac{\pi\alpha's}{4}\sin\varepsilon\right)}\,.
\end{align}
\end{subequations}
It is now clear that the first term on the r.h.s.~of eq.~\eqref{eq:stringdisc1} is responsible for the reproduction of the $t$-channel pole as is discussed in the previous section \ref{correcttreat}, and the remaining pieces give manifestly finite contributions. The first term can be parameterized in the language of the parametrization of Regge behavior \eqref{regge3} as
\begin{align}
f=\frac{256A}{\alpha'^4}\,,\quad f'=\frac{128A\gamma}{\alpha'^3}\,,\quad j'=\frac{\alpha'}{2}\,,\quad j''=0\,,\label{eq:corresp1}
\end{align}
where $\gamma$ is the Euler's constant.
From eqs.~\eqref{eq:smooth1} and \eqref{eq:defdiv2}, $D\,\bigl(t,M_*^2\bigr)$ can be expressed as 
\begin{align}
D\left(t;M_*^2\right)\approx \frac{16A}{{\alpha'}^{2}}\left[-\frac{2}{\alpha't}-\ln\left(\frac{\alpha'M_*^2}{4}\right)-\gamma-\int^\infty_{M_*^2}\mathrm{d}\mu\,\mu^{-1}I(\mu,\varepsilon)-\int^\varepsilon_0\mathrm{d}\theta\,R(M_*^2,\theta)+\mathcal{O}(t)\right]\,.\label{eq:divint1}
\end{align}
The divergent piece in the limit $t\to0$ agrees with eq.~\eqref{eq:singular1}. 
Note that the finite pieces of \eqref{eq:divint1} does not precisely coincide with the one obtained by substituting \eqref{eq:corresp1} into \eqref{uvint2}, but this is simply because we performed the ``smoothing'' of the amplitude by deforming the integration contour as fig.~\ref{smearing}. 

Finally, to check the consistency of the result \eqref{eq:divint1} with \eqref{eq:zero1}, we compute the finite pieces of \eqref{eq:divint1}. For analytic calculations, we choose $\varepsilon$ to satisfy $\alpha'M_*^2\varepsilon^2\ll1$ and $\exp\,\Bigl[\alpha'M_*^2\varepsilon\Bigr]\gg1$ simultaneously. These conditions are compatible with each other because $M_*^2$ is sufficiently larger than ${\alpha'}^{-1}$. With this choice of $\varepsilon$, the fourth term on the r.h.s.~of eq.~\eqref{eq:divint1} can be evaluated as 
\begin{align}
&\int^\infty_{M_*^2}\mathrm{d}\mu\,\mu^{-1}I(\mu,\varepsilon)=-2\int^\infty_{M_*^2}\mathrm{d}\mu\,\frac{2+\cos\left(\frac{\pi\alpha'\mu}{2}\right)}{\mu}\,\exp\left[-\frac{\pi\alpha'\mu\varepsilon}{2}\right]+\mathcal{O}\left(e^{-\alpha'M_*^2\varepsilon}\right)
=\mathcal{O}\left(e^{-\alpha'M_*^2\varepsilon}\right)\,.\label{eq:estimate1a}
\end{align}
When computing the fifth term on the r.h.s.~of eq.~\eqref{eq:divint1}, it is convenient to set $\alpha'M_*^2=4N+2$
 with a sufficiently large integer $N$. In this case, $R(M_*^2,\theta)$ can be expanded in terms of $\theta\ll1$ as\footnote{Note that this expansion is valid under the condition $\alpha'M_*^2\theta^2\ll1$.}
\begin{align}
\left.R\left(M_*^2,\theta\right)\right|_{M_*^2\alpha'=4N+2}=\frac{(2N+1)\pi e^{-(2N+1)\pi\theta}\theta^2}{\left(e^{-(2N+1)\pi\theta}+1\right)^2}+\mathcal{O}\left(\theta^4\right)\,.
\end{align}
 Using this expression, we obtain
\begin{align}
&\int^\varepsilon_0\mathrm{d}\theta\,R(M_*^2,\theta)=\frac{1}{6\left(2N+1\right)^2}+\mathcal{O}\left(N^{-4},\,e^{-N\varepsilon}\right)\,.\label{eq:estimate1b}
\end{align}
Note that eqs.~\eqref{eq:estimate1a} and \eqref{eq:estimate1b} are consistent with the $\varepsilon$-independence of $D\,\bigl(t;M_*^2\bigr)$ at least within the range of our approximation, as it should be. Therefore, we have 
\begin{align}
D\left(t;M_*^2\right)\approx \frac{16A}{{\alpha'}^{2}}\left[-\frac{2}{\alpha't}-\ln\left(N+\frac{1}{2}\right)-\gamma-\frac{1}{6\left(2N+1\right)^2}+\mathcal{O}\left(N^{-4},\,e^{-N\varepsilon}\right)+\mathcal{O}\left(t\right)\right]\,,
\end{align}
leading to\footnote{This behavior \eqref{eq:zero2} completely agrees with the equality for the $N$-th partial sum of the harmonic series
\begin{align}
\sum_{J=1}^N\frac{1}{J}=\ln N+\gamma+c_N\,,
\end{align}
where $c_N$ denotes a constant which vanishes in the limit $N\to\infty$.}
\begin{align}
B^{(2,0)}_{\lth}(0)&\approx \frac{64A}{\pi\alpha'^2}\left[\sum_{J=1}^N\frac{1}{J}-\ln\left(N+\frac{1}{2}\right)-\gamma-\frac{1}{6\left(2N+1\right)^2}+\mathcal{O}\left(N^{-4},\,e^{-N\varepsilon}\right)\right]
=\mathcal{O}\left(N^{-4}\right)\,,\label{eq:zero2}
\end{align}
which coincides with the result \eqref{eq:zero1} within the range of our approximation. 
In this example, $D\,\bigl(t;M_*^2\bigr)$ gives the {\it negative} corrections to $B^{(2,0)}_{\lth}(0)$ which leads to $B^{(2,0)}_{\lth}(0)=0$. This is consistent with what we obtained in the previous sec.~\ref{correcttreat}. 

\subsection{Bound on negativity and the approximate positivity}\label{appposi}
As we have argued in sec.~\ref{correcttreat}, the violation of strict positivity $B^{(2,0)}_{\lth}(0)>0$ is due to the negative corrections to $B^{(2,0)}_{\lth}(0)$ from the residuals of the $t$-channel pole cancellation on the r.h.s.~of eq.~\eqref{subtdisp8}. We have checked in sec.~\ref{subsec:stringeg} that the negative corrections indeed play a crucial role to get the amplitude with vanishing $s^2$ term when performing the low-energy expansion. Because we identified the form of the negative term, we can put an upper bound on the allowed amount of the positivity violation. When $B^{(2,0)}_{\lth}(0)<0$, $-B^{(2,0)}_{\lth}(0)$ is bounded from above,
\begin{align}
-B^{(2,0)}_{\lth}(0)<\frac{f\alpha'^2}{4\pi}\left[\frac{f'}{fj'}+\ln\left(\frac{\alpha'M^2_*}{4}\right)-\frac{j''}{2(j')^2}\right]\,.\label{eq:negabd}
\end{align}
From eq.~\eqref{tpolecancel2}, we find that this upper bound is suppressed by $M_{\rm pl}^{-2}$. The violation of the positivity of order of $\Mpl^{-2}$ is allowed. This  violation will be also acceptable purely from low-energy EFT perspective \cite{deRham:2020zyh}. Specifically, 
assuming that $j'\sim(f'/f)\sim~(j''/j')\sim\alpha'$, we can estimate the r.h.s.~of \eqref{eq:negabd} as
\begin{align}
\frac{f\alpha'^2}{4\pi}\left[\frac{f'}{fj'}+\ln\left(\frac{\alpha'M^2_*}{4}\right)-\frac{j''}{2(j')^2}\right]\sim \frac{\alpha'}{M_{\rm pl}^2}\,.\label{eq:negaest}
\end{align}
In string theory examples, these assumptions are satisfied and $\alpha'=M_{\rm s}^{-2}$ as we saw in sec.~\ref{subsec:stringeg} (although there we have $j''/j'=0$). We expect that the bound on negativity \eqref{eq:negabd} might be improved by requiring more precise knowledge of UV completion, which may allow us to compute the first term of the r.h.s.~of \eqref{subtdisp8}. We leave this aspect for future work. 

An uncertainty of the sign of the residuals of the $t$-channel pole cancellation is not problematic when the r.h.s.~of eq.~\eqref{subtdisp8} is dominated by the first term:
\begin{align}
B_{\lth}^{(2,0)}\left(0\right)\simeq\frac{4}{\pi}\int^{M_*^2}_{\lth^2}\mathrm{d}\mu\,\frac{\im\, F\left(\mu,-0\right)}{\left(\mu-2m_\phi^2\right)^{3}}&\gg\left|\frac{f\alpha'^2}{4\pi}\left[\frac{f'}{fj'}+\ln\left(\frac{\alpha'M^2_*}{4}\right)-\frac{j''}{2(j')^2}\right]\right|
\,. \label{validcondition}
\end{align}
Using the estimation \eqref{eq:negaest}, the condition \eqref{validcondition} reduces to 
\begin{align}
B_{\lth}^{(2,0)}(0)\simeq\frac{4}{\pi}\int^{M_*^2}_{\lth^2}\mathrm{d}\mu\,\frac{\im\, F\left(\mu,-0\right)}{\left(\mu-2m_\phi^2\right)^{3}}&\gg \mathcal{O}\left(\frac{\alpha'}{M_{\rm pl}^2}\right)\,.\label{approxpositivity}
\end{align}
The condition \eqref{approxpositivity} will be satisfied when the cutoff scale of EFT is much lower than the typical energy scale of UV completion $\alpha'^{-1/2}$ or $M_{\rm pl}$. This situation will be realized (i) when the light field $\phi$ is coupled directly to the heavy state whose mass is much lighter than $\bigl(M_{\rm pl}^2/\alpha'\bigr)^{1/4}$ , (ii) or when the light field $\phi$ is coupled to the heavy state only gravitationally whose mass is much lighter than ${\alpha'}^{-1/2}$, for instance.
This gives an {\it approximate} lowest-order positivity bound which is useful to constrain EFT parameters. 
For example, let us consider the simple model of the form 
\begin{align}
S=\int\mathrm{d}^4x\sqrt{-g}\,\left[\frac{M_{\rm pl}^2}{2}R-\frac{1}{2}\left(\der\phi\right)^2-\frac{m_\phi^2}{2}\phi^2+\frac{c}{\Lambda^4}\left(\der\phi\right)^4+({\rm higher\,\,order\,\, terms \,\,in}\,\,\, \Lambda)\right]\,.\label{example1}
\end{align}
The lowest-order approximate positivity imposes $c>0$ within an accuracy of $\mathcal{O}\,\Bigl(\bigl(\Lambda^4/{\alpha'}^{-1}M_{\rm pl}^2\bigr)\Bigr)$. This means that as long as one discusses the positivity bounds on the derivative couplings which play an important role at a center-of-mass energy scale $E\sim\Lambda\ll\bigl(M_{\rm pl}^2/\alpha'\bigr)^{1/4}$, the bound $c\gtrsim0$ indeed works well as a constraint on EFT. In string theory, typically we have ${\alpha'}^{-1/2}\sim 10^{16} \,\,{\rm GeV}$. Therefore we expect that the approximate positivity bounds can be safely applied to modified gravity models where the cutoff scale is usually supposed to be much lower than  $10^{16}\,\,{\rm GeV}$. Especially we apply positivity bounds on scalar-tensor theories of gravity in sec.~\ref{application} and \ref{sec:pheno}. Note that the discussion here can be extended to the approximate positivity bounds with $t$ derivatives in a straightforward manner: see appendix~\ref{tderivext2}.

The condition \eqref{approxpositivity} simply says that the subtlety of lowest-order positivity bounds can be negligible when the contributions from the Regge states can be ignored. This argument has been already made in the appendix of \cite{Hamada:2018dde}. They however did not explicitly evaluate the $t$-channel pole cancellation and did not identify the finite contribution. On the other hand we identified possible negative corrections to $B^{(2,0)}_{\lth}(0)$, and obtained the upper bound on the modulus of the negativity of $B^{(2,0)}_{\lth}(0)$.

\section{Bounds on scalar-tensor EFT}\label{application}
In this section, we apply the approximate positivity bounds to scalar-tensor EFT. The scalar-tensor EFT involves two kinds of operators, non-minimal couplings to gravity such as $G^{\mu\nu}\der_\mu\phi\der_\nu\phi$ and self-interactions of the scalar as discussed in \eqref{example1}. We shall use $M$ to denote the mass scale of the non-minimal couplings whereas $\Lambda$ is referred to as the mass scale of the self-interactions. To obtain meaningful bounds on the EFT, we need to suppose the mass scales are sufficiently lower than the UV completion scale of gravity, namely $M^2 \ll \alpha'{}^{-1}$ and $\Lambda^4 \ll \Mpl^2 \alpha'{}^{-1}$. \emph{This is an inevitable assumption to derive the positivity bounds on scalar-tensor theories without the subtleties.} 

In sec.~\ref{subsec:frame}, we briefly discuss how our approximate positivity bounds work in different frames. In sec.~\ref{subsec:EFTpoint}, we briefly explain the general strategy of our computation. In sec.~\ref{subsec:action}, we derive the most general action for scalar-tensor EFT up to mass dimension 12, by utilizing the freedom of perturbative field-redefinitions and eliminating the terms which are unnecessary when discussing the $2$ to $2$ scattering of $\phi$. We then present positivity bounds on scalar-tensor EFT in sec.~\ref{subsec:EFTbound}.

\subsection{Frame-independence: example}\label{subsec:frame}
In the example \eqref{example1}, we have seen that the approximate positivity provides a meaningful bound on the EFT as far as the scale $\Lambda$ in \eqref{example1} satisfies $\Lambda^4 \ll \Mpl^{2} \alpha'{}^{-1}$. Let us then consider another example to illustrate the efficacy of the approximate positivity in the presence of non-minimal coupling to gravity.
We study the tree-level amplitude of the theory
\begin{align}
S=\int d^4x \sqrt{-g}\left[\frac{M_{\rm pl}^2}{2}R -\frac{1}{2} (\partial \phi)^2 -\frac{m_{\phi}^2}{2} \phi^2 +d_1 R\phi^2 
+\frac{d_2}{M^2}G^{\mu\nu} \partial_{\mu}\phi \partial_{\nu}\phi 
\right]\,, \label{Jordan}
\end{align}
where $R$ and $G^{\mu\nu}$ are the Ricci scalar and the Einstein tensor, respectively, and $M$ determines the mass scale of the non-minimal coupling.
Contrary to the previous example \eqref{example1}, the scalar field $\phi$ has no self-interaction but is non-minimally coupled to gravity.
We shall call \eqref{Jordan} the action in {\it Jordan frame}.

In the Jordan frame, $\phi$ is scattered via exchanging the graviton through the minimal and the non-minimal couplings at the tree-level but not via self-interactions.
The pole subtracted scattering amplitude of the process $\phi\phi \rightarrow \phi\phi$ is computed as
\begin{align}
\tilde{B}_{\rm tree}(v,t)=\frac{d_2}{M_{\rm pl}^2 M^2} \left[2(1-4d_1)-3d_2 \frac{t}{M^2} \right]v^2+O(v)\,,
\label{B_Jordan}
\end{align} 
where $v$ is defined by $s=v+2m_\phi^2-(t/2)$. The terms $O(v)$ includes terms linear in $v$ and independent of $v$ which are irrelevant to the positivity bounds.
To obtain a meaningful bound from the approximate positivity, we need to suppose 
\begin{align}
M^2 \ll \alpha'{}^{-1}\,,
\end{align}
which is different from the previous assumption $\Lambda^4\ll \Mpl^2\alpha'{}^{-1}$ in \eqref{example1} because we have the Planck suppression in \eqref{B_Jordan} due to the gravitational origin of the interactions.
Then, the leading positivity bound $B^{(2,0)}\gtrsim 0$ yields a constraint
\begin{align}
d_2(1-4d_1)\gtrsim 0\,,
\end{align}
whereas the bound $B^{(2,1)}+\frac{3}{2\lth^2}B^{(2,0)}\gtrsim 0$, which is derived in the appendix.~\ref{tderivext} (see eq.~\eqref{abound21}), gives the upper bound on the cutoff scale
\begin{align}
\frac{M^2}{\lth^2}\gtrsim \frac{d_2}{1-4d_1}
\,.
\end{align}
It would be worth noting that the $v^2$ part of \eqref{B_Jordan} vanishes when $d_2=0$, i.e.~the contribution from the minimal coupling is subtracted at $v^2$ (see below as well). The positivity bounds provide a constraint on the sign of a combination of the non-minimal couplings (see also \cite{deRham:2019ctd}).

On the other hand, one can redefine the metric to eliminate the non-minimal couplings. The couplings $R\phi^2$ and $G^{\mu\nu}\partial_{\mu}\phi\partial_{\nu}\phi$ are eliminated by changing the metric as
\begin{align}
g_{\mu\nu}\to Ag_{\mu\nu}+B\partial_{\mu}\phi \partial_{\nu}\phi
\label{disformal_trans}
\end{align}
with
\begin{align}
A&=1-\frac{2d_1}{\Mpl^2}\phi^2 + O(\phi^4)
\,, \\
B&=\frac{2d_2}{\Mpl^2 M^2}+ O(\phi^2)
\,.
\end{align}
Up to here the transformation is invertible even when the fields are not treated as perturbations. 
To minimize the number of operators, we further perform the perturbative field redefinition 
\begin{align}
\phi \to \phi - \frac{2d_1(12d_1-1)M^2+d_2(12d_1+1)\mass^2}{6\Mpl^2 M^2} \phi^3-\frac{12d_1d_2}{\Mpl^2 M^2}X\phi+O(\phi^5)\,,
\label{change_phi}
\end{align}
where $X=-\frac{1}{2}g^{\mu\nu}\partial_{\mu}\phi \partial_{\nu}\phi$. The field redefinition \eqref{change_phi} is invertible only when $\phi$ is treated as a perturbation, differently from \eqref{disformal_trans}.
The action is then
\begin{align}
S=\int d^4x \sqrt{-g}\, \Biggl[ &\frac{M_{\rm pl}^2}{2}R - \frac{1}{2}(\partial \phi)^2 -\frac{1}{2}m_{\phi}^2 \phi^2 
- \left( -\frac{d_1(12d_1+5) \mass^2}{3\Mpl^2} - \frac{d_2(12d_1+1) \mass^4}{6\Mpl^2 M^2} \right)\phi^4
\no\\
&+\frac{2d_2(1-4d_1)}{\Mpl^2 M^2}X^2
+\frac{2d_2^2}{\Mpl^2 M^4}X \left[ (\Box \phi)^2-\phi_{\mu\nu}^2 \right]+\cdots
\Biggl]
\label{Einstein_frame}
\end{align}
where $\phi_{\mu\nu}^2=\nabla_{\mu}\nabla_{\nu}\phi \nabla^{\mu}\nabla^{\nu}\phi$ and $\cdots$ are the terms which are irrelevant to the $\phi\phi\to\phi\phi$ scattering at the tree level. We refer to \eqref{Einstein_frame} as the action in {\it Einstein frame}. In \eqref{Einstein_frame}, the scalar field $\phi$ is minimally coupled with the metric up to the quadratic order in $\phi$ while it has self-interactions. The self-interactions are suppressed by particular combinations of $\Mpl$ and $M$ due to their gravitational origin. The tree-level scattering amplitude of $\phi\phi \rightarrow \phi \phi$ can be divided into two parts
\begin{align}
F_{\rm tree}(s,t)=F_{\rm min}(s,t)+F_{\rm scalar}(s,t)
\end{align}
where $F_{\rm min}$ is the contribution from the graviton exchange through the minimal coupling and $F_{\rm scalar}$ is from the purely scalar scattering. The pole subtracted amplitude of $F_{\rm min}$, denoted by $\tilde{B}_{\rm min}(v,t)$, is
\begin{align}
\tilde{B}_{\rm min}=\frac{1}{2M_{\rm pl}^2}(8m_{\phi}^2-3t-2v)\,,
\end{align}
and thus it does not contribute to the positivity bounds. The non-zero contributions to the positivity bounds are obtained from the self-interaction part of the scattering amplitude
\begin{align}
\tilde{B}_{\rm scalar}=\frac{d_2}{M_{\rm pl}^2 M^2} \left[2(1-4d_1)-3d_2 \frac{t}{M^2} \right]v^2+O(v)\,.
\end{align}
Needless to say, the result obtained in the Einstein frame is the same as that in the Jordan frame.

In the following subsection, we shall derive positivity bounds on general scalar-tensor EFT Lagrangian.
The consideration on the example Lagrangian \eqref{Jordan} explicitly clarifies the following points: (i) the coupling constants of the non-minimal couplings are constrained by the approximate positivity when $M^2 \ll \alpha'{}^{-1}$; (ii) the bounds are independent of field redefinitions as they should be; (iii) the graviton exchange does not contribute to the bounds when the scalar field is minimally coupled to gravity. In particular, the last point would be practically useful for discussing the bounds on the general scalar-tensor EFT. The effect of gravity on the positivity bounds can be simply ignored in the Einstein frame at the tree level calculations. The positivity bounds in the pure scalar system, e.g.~\cite{deRham:2017imi}, can be directly applied to the case in the presence of gravity if the scalar field is minimally coupled to gravity, as long as its cutoff is much lower than the Reggeization scale.

\subsection{Ghost-freeness v.s. EFT point of view}\label{subsec:EFTpoint}

Phenomenologically ghost-free scalar-tensor theories have been discussed, and positivity bounds on a certain subclass of such theories called Horndeski theories have been also discussed in \cite{Melville:2019wyy}. The ghost-freeness is required in order that an unwanted degree of freedom does not appear even at a non-perturbative regime. In a standard local QFT, however, one has to add all the non-renormalizable terms respecting symmetries under consideration when considering the model with non-renormalizable terms. It is necessary to take into account ghostly operators generically speaking, unless one adopts fine-tuning or invokes some physical mechanisms such as symmetrical reasonings which ensures the absence of ghostly operators.\footnote{For example, in Galileon models, it is technically natural to tune the coefficients of ghost-free terms thanks to the non-renormalization theorem, but it is not the case for ghostly operators which receive $\mathcal{O}(1)$ renormalizations.}
We thus consider a scalar-tensor EFT, instead of a ghost-free scalar-tensor theory, which includes all possible operators. 

As we have discussed above, it is well-known that the Lagrangian description contains redundancies associated with field redefinitions. In the context of ghost-free scalar tensor theories, the disformal transformation \eqref{disformal_trans} has been used to connect a theory to another equivalent theory where the equivalence is guaranteed when theories are related by an invertible transformation~\cite{Takahashi:2017zgr}. On the other hand, in the context of the scalar-tensor EFT, one can also perform perturbative field redefinitions like \eqref{change_phi} since the operators are treated as perturbations.
Then, the terms proportional to $R_{\mu\nu}$ or $R$ can be eliminated perturbatively by redefining graviton fluctuations $h_{\mu\nu}$, and the terms proportional to $\Box\phi$ can be eliminated by redefining a scalar field $\phi$ (see for instance, \cite{Solomon:2017nlh}). Let us detail this point in the next subsection.

\subsection{General action}\label{subsec:action}
Let us derive the most general EFT Lagrangian which contains terms contributing to the $\phi\phi\to\phi\phi$ scatterings up to $\mathcal{O}(\Mpl^{-2}M^{-6})$ and $\mathcal{O}(\Lambda^{-8})$, i.e.~up to mass-dimension 12 in the Lagrangian, by utilizing perturbative field redefinitions. We ignore contributions of order of $\mathcal{O}(\Mpl^{-4})$, namely the graviton loops. Note that it has been already studied in previous works how much one can simplify the interaction terms in scalar-tensor theories with a shift symmetry of $\phi$: up to terms containing 6 derivatives in \cite{Solomon:2017nlh} and 8 derivatives in \cite{Ruhdorfer:2019qmk}.\footnote{Regarding the scalar self-interactions, \cite{Solomon:2017nlh} discussed operators up to mass-dimension 12.} In this study, on the other hand, we do not assume a shift symmetry first and 
take into account all operators that contribute to the $\phi\phi\to\phi\phi$ scattering amplitudes up to $\mathcal{O}(\Mpl^{-2}M^{-6})$ and $\mathcal{O}(\Lambda^{-8})$.

Firstly we discuss how much we can simplify the terms with curvatures by making use of field redefinitions.
Especially it is important to investigate whether one has to add operators which contain Riemann tensors ({\it e.g.}, $R_{\mu\nu\rho\sigma}\Box R^{\mu\nu\rho\sigma}$) when considering the $\phi\phi\to\phi\phi$ scattering, because such operators cannot be eliminated perturbatively by any field redefinition. 
Throughout this paper we have assumed that graviton loops can be neglected. Under this assumption, relevant operators to the $\phi\phi\to \phi\phi$ scattering take the following schematic forms
\begin{subequations}
\label{vertexlist1}
\begin{align}
&{ 
{\rm curvature-only\,\, terms}:}\quad ({\rm curv})\,\nabla^{m}({\rm curv})\,,\label{vertexlist1a}\\
&{ 
{\rm scalar-curvature \,\,couplings}:}\quad({\rm curv})\,\nabla^m\phi\,,\quad({\rm curv})\,(\nabla^m\phi)(\nabla^{m'}\phi)\,.\label{vertexlist1b}
\end{align}
\end{subequations}
Here, we symbolically refer to $R$, $R_{\mu\nu}$, and $R_{\mu\nu\rho\sigma}$ as $({\rm curv})$, and the symbol $\nabla^m$ expresses $m$-th order covariant derivatives with indices omitted. By the use of the equivalence upon integration by parts, we can assume that the derivatives of \eqref{vertexlist1a} only act on one of the curvature and those of \eqref{vertexlist1b} act on $\phi$ without loss of generality. Terms of the form \eqref{vertexlist1} contribute to the graviton propagator, graviton-scalar 2-point vertexes, and  graviton-scalar-scalar 3-point vertexes, simply denoted by $hh,h\phi$, and $h\phi\phi$, at tree-level, respectively; they can be relevant to the $\phi\phi\to\phi\phi$ scattering even when neglecting graviton loops. It should be noted that we do not need to take care of the ordering of covariant derivatives when considering operators of the form \eqref{vertexlist1}, because they commute with each other up to the curvature leading to higher order interactions, $hhh, hh\phi$ or $hh\phi\phi$. They are irrelevant to the $\phi\phi\to\phi\phi$ scattering when graviton loops are neglected. 

From now on, we prove that we do not need to include the Riemann tensor as an independent ingredient for our consideration. First, we consider the operators of the form \eqref{vertexlist1a} which change the graviton propagator.
The Riemann squared term $R_{\mu\nu\rho\sigma}R^{\mu\nu\rho\sigma}$ is eliminated by adding the Gauss-Bonnet term which does not affect local physics in 4-dimensions. For $m>0$, the Lorentz invariance concludes that the indices of covariant derivatives must be contracted with either the one of Riemann tensor or of other covariant derivatives, say $R_{\mu\nu\rho\sigma}\Box R^{\mu\nu\rho\sigma}$ and $R^{\nu\rho}\nabla^\mu\nabla^\sigma R_{\mu\nu\rho\sigma}$.  Therefore, neglecting the ordering of the covariant derivatives, the curvature-only terms with $m\neq0$ in eq.~\eqref{vertexlist1a} are always proportional to $\Box R_{\mu\nu\rho\sigma}$ or $\nabla^\mu R_{\mu\nu\rho\sigma}$. The covariant derivatives of the Riemann tensor can be re-expressed in terms of $R$, $R_{\mu\nu}$ and their covariant derivatives up to linear in the curvature, thanks to the following identities which are derived from the Bianchi identities:
\begin{subequations}
\label{bianchi1}
\begin{align}
&\nabla^\mu R_{\mu\nu\rho\sigma}=2\nabla_{[\rho}R_{\sigma]\nu}\,,\label{bianchi1a}\\
&\Box R_{\mu\nu\rho\sigma}=2\left(\nabla_\rho\nabla_{[\mu}R_{\nu]\sigma}-\nabla_\sigma\nabla_{[\mu}R_{\nu]\rho}\right)+({\rm curv})^2\,.\label{bianchi1b}
\end{align}
\end{subequations}
Here,  
the symbol $[\cdots]$ denotes a standard anti-symmetrization, $2A_{[\mu\nu]}\coloneqq (A_{\mu\nu}-A_{\nu\mu})$. Hence,  operators, $({\rm curv})\,\nabla^{m}({\rm curv})$, are proportional to either the Ricci tensor or the Ricci scalar.

Next, let us consider operators \eqref{vertexlist1b} where the former ones produce mixing of the graviton and the scalar field, $h\phi$, while the latter ones yield 3-point vertexes, $h\phi\phi$. As for the first type, $h\phi$, it is obvious that operators of the form $R^{\mu\nu\rho\sigma} \nabla^m \phi$ are irrelevant to our consideration due to the anti-symmetric property of the Riemann tensor. For instance, the lowest-order term with the Riemann tensor is $R_{\mu\nu\rho\sigma}\nabla^{[\mu}\nabla^{\nu]}\nabla^{[\rho}\nabla^{\sigma]}\phi$ which contributes from a 4-point vertex $hhh \phi$ since the commutation of the covariant derivatives can be replaced with the curvature. The second type, $h\phi\phi$, may take the form of $R^{\mu\nu\rho\sigma}(\nabla_{\mu}\nabla_{\rho} \nabla^{m-2}\phi)(\nabla_{\nu}\nabla_{\sigma} \nabla^{m'-2}\phi)$ up to the ordering of covariant derivatives. In particular, we have  a ghostly operator $R^{\mu\nu\rho\sigma}\nabla_{\mu}\nabla_{\rho}\phi \nabla_{\nu}\nabla_{\sigma}\phi$ which cannot be eliminated by any field redefinition~\cite{Solomon:2017nlh}.
However, the contribution of the term $R^{\mu\nu\rho\sigma}(\nabla_{\mu}\nabla_{\rho} \nabla^{m-2}\phi)(\nabla_{\nu}\nabla_{\sigma} \nabla^{m'-2}\phi)$ to $\phi\phi\to\phi\phi$ scatterings via tree-level graviton exchange is redundant since
\begin{align}
&R_{\mu\nu\rho\sigma}(\nabla^\mu\nabla^\rho \nabla^{m-2}\phi)(\nabla^\nu\nabla^\sigma \nabla^{m'-2}\phi)\no\\
&=-2\nabla_{[\rho}R_{\sigma]\nu}(\nabla^\rho \nabla^{m-2}\phi)(\nabla^\nu\nabla^\sigma \nabla^{m'-2}\phi)-R_{\mu\nu\rho\sigma}(\nabla^\rho \nabla^{m-2}\phi)(\nabla^{[\mu}\nabla^{\nu]}\nabla^\sigma \nabla^{m'-2}\phi)+({\rm total\,\,derivative})\,,
\end{align}
where we used \eqref{bianchi1a}. The first term is further rewritten as the form $R_{\mu\nu}\nabla^n\phi \nabla^{n'}\phi$ via integration by parts. The second term contains Riemann tensor and cannot be eliminated by any field redefinition, but it contains at least 2 curvatures and hence it is irrelevant to our analysis.
As a result, we conclude that operators \eqref{vertexlist1a} and \eqref{vertexlist1b} are proportional to the Ricci tensor/scalar and thus they can be eliminated by a perturbative field redefinition of the metric perturbation.

The most general EFT Lagrangian for scalar-tensor theories can be written as
\begin{align}
\mathcal{L}=\frac{M_{\rm pl}^2}{2}R-\frac{1}{2}(\der\phi)^2-\frac{m_\phi^2}{2}\phi^2+\mathcal{L}_{\rm s}[\phi,\nabla]+RF_1[\phi,\nabla, R, R_{\rho\sigma}]+R_{\mu\nu}F^{\mu\nu}_2[\phi,\nabla, R, R_{\rho\sigma}]+\mathcal{L}_{\rm higher}\,, \label{lag1}
\end{align}
before performing the field redefinitions
where $\lag_{\rm higher}$ contains all higher-order terms which do not contribute to the $\phi\phi\to\phi\phi$ scattering up to mass dimension 12 when the graviton loop corrections are neglected.\footnote{Here, all the operators whose mass-dimension is higher than 12 are contained in $\lag_{\rm higher}$ which we will discard in this study. We assume that the dimensionless coefficients of higher-order operators are not anomalously large so that we can safely discard them in this study.} All the interaction terms which are relevant to our study here are contained in $\lag_{\rm s}+RF_1+R_{\mu\nu}F^{\mu\nu}_2$. $\lag_{\rm s}$ denotes self-interaction terms of $\phi$, namely $\phi\phi\phi$ and $\phi\phi\phi\phi$. $F_1$ and $F^{\mu\nu}_2$ are arbitrary functions consisting of $\phi$,  $R$, $R_{\rho\sigma}$, and covariant derivatives $\nabla$ which provide $hh, h\phi$ and $h\phi\phi$. In principle, we can eliminate $F_1$ and $F_2$ by a perturbative field redefinition. We then obtain the Einstein frame where the scalar field $\phi$ is minimally coupled to gravity up to the relevant interactions to the $\phi\phi \to \phi\phi$ scattering. However, in practice, it would be better to retain the ghost-free operators and to derive the bounds on the coupling constants of them since they have been widely discussed for phenomenological purposes. After appropriately performing the field redefinitions to eliminate  ghostly operators of $F_1$ and $F_2$, we obtain the following Lagrangian:
\begin{align}
\mathcal{L'}&=\frac{M_{\rm pl}^2}{2}R-\frac{1}{2}(\der\phi)^2-\frac{m_\phi^2}{2}\phi^2+\mathcal{L}'_{\rm s}[\phi,\nabla] 
\no \\
&+d_{10}\Mpl R \phi + d_{20} R  \phi^2+\frac{d_{22,1}}{M^2}G^{\mu\nu}(\der_\mu\phi)(\der_\nu\phi)+\frac{d_{22,2}}{M^2}RX+\mathcal{L}'_{\rm higher}\,,
\label{lag2}
\end{align}
with $X=-\frac{1}{2}(\partial \phi)^2$. The first two indices of the coupling constants $d_{ij,k}$ represent the number of $\phi$ and of derivatives while the last index $k$ classifies different operators if there exists several operators with the same $i$ and $j$. There are four ghost-free non-minimal couplings $R\phi, R\phi^2,G^{\mu\nu}\partial_{\mu}\phi \partial_{\nu}\phi$ and $RX$.
Note that although the term $RX$ yields higher order equations of motion, one can add a counter term to $\mathcal{L}_s'$ of the form $(\partial X)^2$. The equations of motion are then degenerate and free from the Ostrogradsky ghost~\cite{Langlois:2015cwa}. We thus retain the coupling $RX$ on an equal footing with the other non-minimal couplings. The coupling $R\phi$ yields a mixing between the graviton fluctuations and the scalar fluctuations. Although the coupling $R\phi$ can be included in the interactions when $d_{10}\ll 1$, the typical value of $d_{10}$ is of order unity in the context of modified gravity. In this case, one cannot treat $R\phi $ as a perturbation and thus one should perform a field redefinition to eliminate the coupling $R\phi$. In this section, we simply suppose $d_{10}=0$ and regard the Lagrangian \eqref{lag2} as the one after the field redefinition. The discussion on the coupling $R\phi $ is given in Appendix \ref{sec:all_bound}.

We then consider a perturbative field redefinition of $\phi$ to remove the ghostly operators of scalar self-interaction as much as possible. Again, we do not care about the ordering of covariant derivatives acting on $\phi$ because the commutation of the derivatives is replaced with the curvature yielding the interactions, $h\phi,h\phi\phi$ or more higher orders, which are already discussed above. 
The situation here is essentially the same as the pure scalar field theory without gravity. In this setup, independent operators have been already identified in previous studies: up to mass dimension-$10$ operators in \cite{deRham:2017imi} in the absence of a shift symmetry of $\phi$, and up to mass dimension-$12$ operators in \cite{Solomon:2017nlh} in a shift-symmetric model.\footnote{Note that higher-point interaction vertexes higher than 4 are neglected in \cite{deRham:2017imi}.} By writing down all the operators up to mass dimension $12$ and eliminating ghostly operators when they are proportional to $\Box\phi$, it is straightforward to find that the most generic form  of the scalar part\footnote{We do not consider higher-point interaction vertexes such as $X^3$ because they contribute to $F(s,t)$ only at loop level and can be ignored.}
\begin{align}
\lag'_{\rm s}&=c_{30}m_{\phi} \phi^3+c_{40}\phi^4+\frac{c_{32}}{\Lambda}\phi X+\frac{c_{42}}{\Lambda^2}\phi^2X
+\frac{g_3}{\Lambda^3}X\Box\phi
+\frac{c_{44,1}}{\Lambda^4}X^2
+\frac{c_{44,2}}{\Lambda^4}\phi X \Box \phi
\no \\
&+\frac{g_4}{\Lambda^6}X\left[(\Box\phi)^2-\phi_{\mu\nu}^2\right]
+\frac{3d_{22,2}^2}{\Mpl^2M^4} (\partial_{\mu}X)^2
+\frac{c_{48}}{\Lambda^8}\left(\phi_{\mu\nu}^2\right)^2\,,
 \label{scalarlag}
\end{align}
where the terms $g_3$ and $g_4$ are particularly called the cubic Galileon and the quartic Galileon, respectively. The term $(\partial X)^2$ is introduced so that the equations of motion is degenerate; thus, the coefficient is fixed by that of the coupling $RX$.
On the other hand, the term $g_4$ also requires the counter term $X^2R$ to make it ghost-free. However, the term $X^2R$ is a 5-point interaction $h\phi\phi\phi\phi$ which is irrelevant to our consideration.

Interestingly, one can write the general EFT Lagrangian in a ghost-free form up to mass-dimension 11, while a ghostly operator $(\phi_{\mu\nu}^2)^2$ appears at mass-dimension 12 that cannot be eliminated by any field redefinition. Note that the appearance of this ghostly operator at mass-dimension 12 in scalar self-interactions was already found in \cite{Solomon:2017nlh} in a shift-symmetric case.

\subsection{Positivity bounds on scalar-tensor EFT}\label{subsec:EFTbound}
We shall derive the bounds on the Lagrangian \eqref{lag2} with \eqref{scalarlag} by evaluating scattering amplitudes up to $\mathcal O(\Mpl^{-2}M^{-6})$ and $\mathcal O(\Lambda^{-8})$. In the case of the tree-level calculations, one can straightforwardly obtain the bounds. However, the tree-level approximation may or may not be valid depending of the values of the coupling constants.  For instance, as discussed in \cite{Herrero-Valea:2019hde} (see also Appendix \ref{subsec:loop}), the renormalizable couplings $c_{i0}$ trivialize the positivity bounds since $c_{i0}$ contribute to the leading bound $B^{(2,0)}>0$ at the one loop level unless $c_{i0}$ are extremely tiny. Fortunately, the dominant contribution from $c_{i0}$ can be subtracted by using the improved positivity bound $B^{(2,0)}_{\lth}>0$ as shown in Appendix \ref{subsec:loop}, but $c_{i0}$ still produces a contribution with the factor $\Lambda^4/\lth^4$. The situation becomes more complicated when considering the higher order bounds, e.g. $B^{(4,0)}>0$ which could give a bound on the ghostly coupling $c_{48}$.

To simplify the loop calculations, we further perform the perturbative field redefinition to minimize the number of independent operators. We obtain the Einstein frame action
\begin{align}
\mathcal{L}_E&=\frac{M_{\rm pl}^2}{2}R-\frac{Z}{2}(\partial \phi)^2 -\frac{Z_m m_{\phi}^2}{2} \phi^2
+Y \phi+ \mass Z_{30} c'_{30} \phi^3+Z_{40} c'_{40} \phi^4
\no \\
&+\frac{Z_{44} c'_{44} }{\Lambda^4}X^2
+\frac{Z_{g_4} g_4'}{\Lambda^6}X\left[(\Box\phi)^2-\phi_{\mu\nu}^2\right]+\frac{Z_{48} c'_{48} }{\Lambda^8}\left(\phi_{\mu\nu}^2\right)^2 +{\rm higher~orders}\,,
\label{L_EFT}
\end{align}
where
\begin{align}
c'_{30}&= c_{30}+\frac{\mass}{4\Lambda} \left( c_{32} +\frac{g_3 \mass^2}{\Lambda^2} \right) 
\,, 
\label{c_30} \\
c_{40}'&=c_{40}-\frac{3c_{30} c_{32}\mass }{4\Lambda}-\frac{\mass^2}{\Lambda^2}
\left(\frac{19c_{32}^2}{96}-\frac{c_{42}}{6} \right) -\frac{5c_{30}g_3\mass^3}{4\Lambda^3}
+\frac{\mass^4}{\Lambda^4}\left(\frac{c_{44,2}}{6}-\frac{9c_{32}g_3}{16}\right)
-\frac{9g_3^2 \mass^6}{32\Lambda^6}
\no \\
&+d_{20}(5+12d_{20})\frac{\mass^2}{3\Mpl^2}+\left( \frac{1}{6}d_{22,1}(1+12d_{20})+\frac{1}{3}d_{22,2}(1+6d_{20})\right) \frac{\mass^4}{\Mpl^2M^2}
\,, \\
c_{44}'&=c_{44,1}+c_{32} g_3 + \frac{g_3^2 \mass^2}{2\Lambda^2} +2 \left[ d_{22,1}(1-4d_{20})-d_{22,2}(1+12d_{20}) \right] \frac{ \Lambda^4}{\Mpl^2M^2}
\,, \label{c_44}  \\
g_4'&=g_4-\frac{g_3^2}{2}+2d_{22,1} (d_{22,1}+d_{22,2})\frac{ \Lambda^6}{\Mpl^2M^4}
\,, \\
c_{48}'&=c_{48}
\,, \label{c_48}
\end{align}
and $Z_x$ and $Y$ are introduced to represent the loop corrections to the coupling constants, i.e. $Z_x=1,Y=0$ at the tree-level. At the tree-level, one can easily confirm that the amplitude computed by the original action \eqref{lag2} with \eqref{scalarlag} exactly agrees with that by \eqref{L_EFT} with \eqref{c_30}-\eqref{c_48}. Since the amplitude is invariant under the perturbative field redefinitions, we compute the amplitudes based on the simplified action \eqref{L_EFT} and derive the positivity bounds. By using \eqref{c_30}-\eqref{c_48}, one can straightforwardly translate the bounds on \eqref{L_EFT} to the bounds on \eqref{lag2} with \eqref{scalarlag}.\footnote{The relation of the coupling constants in the presence of the coupling $\phi R$ is shown in Appendix \ref{sec:all_bound}.}

Since $\phi$ is now minimally coupled to gravity in the Lagrangian \eqref{L_EFT}, the discussion on the positivity bound is essentially the same as that in the pure scalar theory~\cite{deRham:2017avq,deRham:2017imi} (see also \cite{Bellazzini:2019xts}). It should be, however, noted that we have two scales $M$ and $\Lambda$ where the former one is associated with the gravitational non-minimal couplings and the latter one is with the self-interactions of higher dimensional operators. In the context of modified gravity, the scalar field $\phi$ is interpreted as a part of gravity, meaning that the scalar self-interactions are supposed to be Planck suppressed. If we postulate that all scalar self-interactions are comparable to the gravitational ones, a natural scaling of the coupling constants would be
\begin{align}
c'_{30}&=O(\mass/\Mpl)\,,\quad
c'_{40}=O(\mass^2/\Mpl^2)\,, 
\no \\
\frac{c'_{44}}{\Lambda^4}&=O(\Mpl^{-2}M^{-2})\,,~
\frac{g_4'}{\Lambda^6}=O(\Mpl^{-2}M^{-4})\,,~
\frac{c'_{48}}{\Lambda^8}=O(\Mpl^{-2}M^{-6})\,,
\label{MG_scaling}
\end{align}
in order that the amplitude is scaled as $\Mpl^{-2}$. 

Hereinafter, we assume an approximate shift symmetry i.e.~the couplings $c'_{i0}$ are suppressed, say $c'_{30}=O(\mass/\Mpl),c'_{40}=O(\mass^2/\Mpl^2)$. This assumption would be reasonable in the context of modified gravity as explained or in the case that $\phi$ is responsible for the acceleration of the universe. In this case, the loop contributions from the renormalizable couplings to the positivity bounds are indeed negligible by the use of the improved positivity bounds. Then, up to $O(\Mpl^{-2}M^{-6})$, the only relevant loop contribution comes from the coupling $c_{44}'$ which contributes to the sub-leading bound $B^{(4,0)}>0$.

We use the dimensional regularization and adopt the $\overline{\rm MS}$-scheme to compute the scalar loops. The detail calculations on the loop are shown in Appendix \ref{sec:Xloop} for the $c_{44}'$ term and Appendix \ref{subsec:loop} for the renormalizable couplings. Due to the approximate shift symmetry, we can ignore the loop contributions from $c'_{i0}$ and then we obtain
\begin{align}
F&=F_{\rm min}+F_{\rm scalar}+F_{\rm 1-loop} + \cdots
\,,
\end{align}
where $F_{\rm min}$ is the contribution from the graviton exchange via the minimal coupling and $\cdots$ are negligible contributions.
The tree amplitude is
\begin{align}
F_{\rm scalar}&=36c'^2_{30}\left[ \frac{1}{\mass^2-s}+\frac{1}{\mass^2-t}+\frac{1}{\mass^2-u}\right]
\no \\
&+24c_{40}' + \frac{c_{44}'}{2\Lambda^4}(s^2+t^2+u^2-4\mass^4) +\frac{3g_4'}{2\Lambda^6}stu 
\no \\
&+ \frac{c_{48}'}{\Lambda^8}\left[\frac{1}{4}(s^2+t^2+u^2)^2-4\mass^2 stu -4\mass^4 (s^2+t^2+u^2)+24 \mass^8 \right]
, 
\end{align}
and the 1-loop amplitude is shown in \eqref{eq:Xloop1}. We then obtain
\begin{align}
B_{\lth}^{(2,0)}&=\frac{2 c_{44}'}{\Lambda^4}+O(c_{44}'^2\mass^2 \lth^2/\Lambda^8)\,, \\
B_{\lth}^{(2,1)}&=-\frac{3g_4'}{\Lambda^6}+O(c_{44}'^2 \mass^2/\Lambda^8)\,, \\
B_{\lth}^{(4,0)}&=\frac{1}{\Lambda^8}\left[24c_{48}'+\frac{3c_{44}'^2\mass^2}{2\pi^2\lth^2}\left(1+O(\mass^2/\lth^2) \right) \right]
.
\end{align}
As for the bounds up to $O(\Lambda^{-6})$, we can ignore the loop corrections. We obtain
\begin{align}
c_{44}' \geq 0\,,  \quad \frac{\Lambda^2}{\lth^2}\geq \frac{g_4'}{c_{44}'}
\,,
\label{leading_bound}
\end{align}
where we include the equality because our leading bounds are approximate ones. The latter bound gives an upper bound on the cutoff $\lth$, typically $\lth \lesssim \Lambda$, when $g_4'> 0$ with $g_4' \sim c_{44}'$. The required cutoff $\lth \lesssim \Lambda$ is the same as the perturbative unitarity cutoff for $g_4',c_{44}'=O(1)$. On the other hand, if we have the scaling \eqref{MG_scaling} with $g_4'>0$, the cutoff $\lth$ is bounded by $M$ where $M$ is not the unitarity cutoff because the amplitude is additionally suppressed by $\Mpl^{-2}$. The bound $\lth \lesssim M$ with $g_4'>0$ may be understood as the requirement of the analyticity rather than the unitarity as discussed in~\cite{deRham:2017avq}. Finally, the second bound of \eqref{leading_bound} does not lead to a bound on $\lth$ if $g_4'\leq 0$. In other words, $g_4'\leq 0$ is required to validate the theory beyond $M$. 

The loop corrections should be included in the sub-leading bound $B^{(4,0)}_{\lth}>0$. The bound is
\begin{align}
c_{48}' > -\frac{c_{44}'^2\mass^2}{16\pi^2\lth^2}\,,
\label{subleading_bound}
\end{align}
where the r.h.s.~is the loop corrections from the $\frac{c_{44}'}{\Lambda^4}X^2$ vertex. The $\mathcal{O}\,\left((\mass^2/\lth^2)\right)$ suppression of the r.h.s.~is due to the  improvement of the positivity bounds \eqref{eq:defimpr}, while the loop corrections to $B^{(4,0)}$ are of order of $\Lambda^{-8}$. Eq.~\eqref{subleading_bound} implies that the absence of the ghostly operator $c_{48}'=c_{48}=0$ would be consistent with the positivity bound owing to the loop corrections, as long as $\mass\neq0$. Interestingly, our result suggests that the ghost-freeness and the sub-leading bound~\eqref{subleading_bound} require the non-zero scalar mass. We should keep in mind, however, that the higher dimensional operators which are neglected in this study might give relevant contributions to the r.h.s. generically, because the r.h.s.~of \eqref{subleading_bound} is suppressed by $\left(\mass^2/\lth^2\right)$. We do not evaluate such additional contributions in this study, but we expect that the sub-leading bound $B^{(4,0)}_{\lth}>0$ would be consistent with the ghost-freeness of scalar-tensor theory with $\mass\neq0$ up to $\mathcal{O}(\Lambda^{-8})$ when the coefficients of higher-dimensional operators are suitably adjusted. 

When the scaling \eqref{MG_scaling} is obeyed, an additional care is required to interpret the result \eqref{subleading_bound}. Under the assumption \eqref{MG_scaling}, the loop correction of $X^2$ is proportional to $\Mpl^{-4}$ that can be thus comparable to graviton loop corrections. This is a trivial consequence because \eqref{MG_scaling} means that the scalar interaction is the same order of magnitude as the gravitational ones. The graviton loops cannot be ignored to evaluate the r.h.s.~of \eqref{subleading_bound}. Nonetheless, the loop contributions are of order of $\mathcal{O}(\Mpl^{-4}M^{-4})$. Hence, even if the graviton loops spoil the consistency of $c_{48}=0$,  
the required bound of the ghostly operator would be
\begin{align}
\frac{c_{48}'}{\Lambda^8} > \mathcal{O}(\Mpl^{-4}M^{-4})
\,.
\end{align}
The ghostly operator can be ignored at $\mathcal{O}(\Mpl^{-2}M^{-6})$. In this sense, accepting the equivalence upon perturbative field redefinitions, the ghost-free scalar-tensor theory would be consistent with the positivity constraints at least up to dimension 12 as long as the mass of a scalar field is non-zero.

\section{Application to phenomenological models}\label{sec:pheno}

In this section, we translate the bounds \eqref{leading_bound} into the language of ghost-free scalar-tensor theories.\footnote{We ignore the loop correction in this section since only the leading bounds \eqref{leading_bound} are relevant for the ghost-free theories.} The previous work~\cite{Melville:2019wyy} discussed the positivity bounds on the Horndeski theory which is the most general scalar-tensor theory with at most second order equations of motion. However, the Horndeski theory is not the most general ghost-free scalar-tensor theory; the known most general theory without the Ostrogradsky ghost is the cubic degenerate higher-order scalar-tensor (DHOST) theory~\cite{BenAchour:2016fzp}. We thus discuss the positivity bounds on the cubic DHOST theory and show how the arbitrary functions of it are constrained.

The most general scalar-tensor Lagrangian involving up to cubic powers of the second derivative of the scalar field $\phi$ is given by
\begin{align}
	{\cal L}_{\rm cD} &= 
	 G_{2}(\phi,X) -G_{3}(\phi,X)\Box\phi 
	  +f_{2}(\phi,X)R
	  +C^{\mu\nu\rho\sigma}_{(2)}\phi_{\mu\nu}\phi_{\rho\sigma}
 +f_{3}(\phi,X)G^{\mu\nu}\phi^{\mu\nu}
	+C^{\mu\nu\rho\sigma\alpha\beta}_{(3)}\phi_{\mu\nu}\phi_{\rho\sigma}\phi_{\alpha\beta}\,,
	\label{classI cD}
\end{align}
where $C^{\mu\nu\rho\sigma}_{(2)}$ and $C^{\mu\nu\rho\sigma\alpha\beta}_{(3)}$ are arbitrary functions constructed by $g_{\mu\nu},\phi$ and $\partial_{\mu}\phi$.
The cubic DHOST is defined such that these functions satisfy the degeneracy conditions that ensure the existence of an additional primary constraint on the theory. We especially focus on the joined class ${}^2$N-I$+ {}^3$N-I in the cubic DHOST~\cite{BenAchour:2016fzp} because the theory is free from instabilities in the cosmological background~\cite{deRham:2016wji,Langlois:2017mxy}. This class of the cubic DHOST includes the Horndeski theory as a subclass. There are 13 conditions on arbitrary functions and then we only have six scalar free functions of $\phi$ and $X$ (see eq.~\eqref{classI relevant cD} below).

In cosmological studies, we usually suppose that the ghost-free terms equally contribute to the background dynamics of the universe. The typical time scale of the universe is the current Hubble scale $H_0$ and the scalar field is supposed to dominate the universe, $\langle \phi \rangle \sim M_{\rm pl}$.
It is then useful to introduce the scales $\Lambda_2= (\Mpl H_0)^{1/2}$ and $\Lambda_3= (\Mpl H_0^2)^{1/3}$.
We define the dimensionless combinations
\begin{align}
	\tilde\phi &:=\frac{\phi}{M_{\rm pl}}, \quad
	\tilde X := -\frac{(\partial\phi)^2}{2\Lambda^4_2}, \quad
	\tilde{\phi}_{\mu}:=\frac{1}{\Lambda_2^2}\partial_{\mu}\phi\,, \quad 
	\tilde{\phi}_{\mu\nu}:=\frac{1}{\Lambda_3^3}\nabla_{\nu}\nabla_{\mu}\phi
\end{align}
and rewrite the Lagrangian as
\begin{align}
	{\cal L}_{\rm cD} &= 
	 \Lambda_{2}^{4}\tilde{G}_{2}(\tilde\phi,\tilde X) -\Lambda_{2}^{4}\tilde{G}_{3}(\tilde\phi,\tilde X)\tilde{\phi}^{\mu}_{\, \mu} 
	 +M_{\rm pl}^{2}\tilde{f}_{2}(\tilde\phi,\tilde X)R
	  -\Lambda^{4}_{2}\tilde{a}_1(\tilde\phi,\tilde X) [(\tilde{\phi}^{\mu}_{\, \mu} )^{2} -\tilde\phi_{\mu\nu}\tilde\phi^{\mu\nu}]
\notag\\
	  &\quad 
	  +\Lambda_{2}^{4}\tilde{a}_3(\tilde\phi,\tilde X) \tilde{\phi}^{\mu}_{\, \mu} \tilde\phi^{\nu}\tilde\phi_{\nu\rho}\tilde\phi^{\rho}
	  +\Lambda_{2}^{4}\tilde{a}_4(\tilde\phi, \tilde X)\tilde\phi^{\mu}\tilde\phi_{\mu\rho}\tilde\phi^{\rho\nu}\tilde\phi_{\nu}
\notag\\
	&\quad +M_{\rm pl}^{2}\tilde{f}_{3}(\tilde\phi,\tilde X)G^{\mu\nu}\tilde{\phi}_{\mu\nu}
	+\Lambda_{2}^{4}\tilde{b}_{1}(\tilde\phi,\tilde X)[(\tilde{\phi}^{\mu}_{\, \mu})^{3}
	-3\tilde{\phi}^{\mu}_{\, \mu}  \tilde\phi_{\nu\rho}^{2}
	+2\tilde\phi_{\mu\nu}^{3}]+\cdots \,,
	\label{classI relevant cD}
\end{align}
by the use of the dimensionless variables and the dimensionless functions $\tilde{G}_i,\tilde{f}_i~(i=2,3),\tilde{a}_1$ and $\tilde{a}_3$
where $\tilde{a}_4$ and $\tilde{b}_1$ are determined by $\tilde{f}_i,\tilde{a}_1,\tilde{a}_3$ due to the degeneracy conditions. In eq. \eqref{classI relevant cD}, we omitted irrelevant terms on the positivity bounds. This normalization implies that we have $\tilde{G}_i,\tilde{f}_i,\tilde{a}_1,\tilde{a}_3 =O(1)$ and $\tilde{\phi},\tilde{X},\tilde{\phi}_{\mu\nu}=O(1)$ around the cosmological background.

The positivity bounds are the bounds on the operators defined around the flat background. We thus suppose that \eqref{classI relevant cD} admits the Minkowski solution under $\phi=$ constant. Without loss of generality, the constant is set to be zero by redefining the origin of $\phi$. Any of arbitrary functions is expanded as Taylor series of $\tilde{\phi}$ and $\tilde{X}$,
\begin{align}
	\tilde{f}(\tilde{\phi}, \tilde{X}) = \bar f + \bar f_{\phi}\tilde{\phi} +\bar f_{X}\tilde{X}
	+\frac{\bar f_{\phi\phi}}{2}\tilde{\phi}^2
	+\bar f_{\phi X} \tilde{\phi} \tilde{X}
	+\frac{\bar f_{XX}}{2}\tilde{X}^2
	+\cdots,
\end{align}
where an overbar indicates the functions evaluated at the flat background, $\phi=0$ and $g_{\mu\nu} =\eta_{\mu\nu}$, and
subscripts $\phi$ and $X$ denote derivatives with respect to $\tilde\phi$ and $\tilde X$ respectively. We have $\bar{G}_{2\phi}=0$ from the requirement that the Minkowski spacetime is a solution. We canonically normalize the field so that $\bar{f}_2=\frac{1}{2}, \bar{G}_{2X}-2\bar{G}_{3\phi}=1$.
The degeneracy conditions read
\begin{align}
\bar{a}_4=-\bar{a}_3+\frac{3(\bar{a}_1+\bar{f}_{2X})^2}{2\bar{f}_2}
\,,\quad
\bar{b}_1=-\frac{1}{6}\bar{f}_{3X}\,, \quad \bar{b}_{1\phi}=-\frac{1}{6}\bar{f}_{3\phi X}\,,
\end{align}
for the relevant coefficients (the full degeneracy conditions are found in \cite{BenAchour:2016fzp}).
After using the degeneracy conditions and performing the integration by parts, we obtain the following relations between the DHOST parameters and the coefficients of \eqref{lag2} and \eqref{scalarlag}:
\begin{align}
\Mpl d_{10}&= \Mpl \bar{f}_{2\phi} \,, \quad
d_{20}= \frac{1}{2}\bar{f}_{2\phi\phi} \,, \quad
\frac{d_{22,1}}{M^2}= -\frac{\bar{a}_1+\bar{f}_{3\phi}}{H_0^2} \,, \quad
\frac{d_{22,2}}{M^2}=\frac{\bar{a}_1+\bar{f}_{2X}}{H_0^2} \,, 
\no \\
\mass c_{30}&= \frac{H_0^2}{6\Mpl} \bar{G}_{2\phi\phi\phi} \,, \quad
\frac{c_{32}}{\Lambda}= \frac{\bar{G}_{2\phi X}-2\bar{G}_{3\phi\phi}}{\Mpl} \,, \quad
\frac{g_3}{\Lambda^3}= -\frac{\bar{G}_{3X}+3\bar{a}_{1\phi}}{\Lambda_3^3} \,, 
\no \\
c_{40}&= \frac{H_0^2\bar{G}_{2\phi\phi\phi\phi} }{24\Mpl^2} \,, \quad
\frac{c_{42}}{\Lambda^2}= \frac{\bar{G}_{2\phi\phi X}-2\bar{G}_{3\phi\phi\phi}}{2\Mpl^2} \,, \quad
\frac{c_{44,1}}{\Lambda^4}= \frac{\bar{G}_{2XX}+4\bar{a}_{1\phi\phi}}{2\Lambda_2^4} \,, 
\no \\
\frac{c_{44,2}}{\Lambda^4}&=-\frac{\bar{G}_{3\phi X}+3\bar{a}_{1\phi\phi}}{\Lambda_2^4} \,, \quad
\frac{g_4}{\Lambda^6}=\frac{3\bar{a}_3-3\bar{a}_{1X}-2\bar{f}_{3\phi X}}{3\Lambda_3^6}
\,,
\end{align}
where the mass of the scalar field $\phi$ is given by $\mass^2=-\bar{G}_{2\phi\phi}H_0^2$. It is now straightforward to obtain the positivity bounds on the DHOST parameters. An important remark is that the coupling constants are scaled as \eqref{MG_scaling} with $M=H_0$. In this case, the second one of \eqref{leading_bound} reads $\lth \lesssim M=H_0$ if $g_4'>0$ but this simply means that the DHOST theory cannot be used for cosmology.
Therefore, the positivity bounds \eqref{leading_bound} should be understood as $c'_{44}\geq 0, -g_4'\geq 0$ in order to consistently describe the cosmological phenomena~\cite{Melville:2019wyy}. In terms of the DHOST parameters, these bounds are given by
\begin{align}
&\frac{1}{2}\bar{G}_{2XX}+2\bar{a}_{1\phi\phi}-(\bar G_{3X} +3\bar a_{1\phi})(\bar G_{2\phi X} -2\bar G_{3\phi\phi})-\frac{1}{2}\bar{G}_{2\phi\phi}(\bar G_{3X} +3\bar a_{1\phi})^{2}
\no \\
-&2(\bar{f}_{2X}+\bar{a}_{1})(6\bar{f}_{2\phi\phi}+1)+2(\bar{f}_{3\phi}+\bar{a}_1)(2\bar{f}_{2\phi\phi}-1) \geq 0
\,, \label{DHOST1} \\
&\bar{a}_{1X}-\bar{a}_3+\frac{2}{3}\bar{f}_{3\phi X}+\frac{1}{2}(\bar{G}_{3X}+3\bar{a}_{1\phi})^2+2(\bar{f}_{2X}-\bar{f}_{3\phi})(\bar{f}_{3\phi}+\bar{a}_1) \geq 0
\,, \label{DHOST2}
\end{align}
where we have assumed $d_{10}=\bar{f}_{2\phi}=0$ for simplicity.
The Horndeski theory is the case $f_2=G_4,f_3=G_5,a_1=-G_{4X},a_3=0$. The bounds are then
\begin{align}
&\frac{1}{2}\bar{G}_{2XX}-2\bar{G}_{4X\phi\phi}-(\bar G_{3X} -3\bar G_{4\phi X})(\bar G_{2\phi X} -2\bar G_{3\phi\phi})-\frac{1}{2}\bar{G}_{2\phi\phi}(\bar G_{3X} -3\bar G_{4\phi X})^{2}
\no \\
&+2(\bar{G}_{5\phi}-\bar{G}_{4X})(2\bar{G}_{4\phi\phi }-1) \geq 0
\,, \label{Horn1}\\
&-\bar{G}_{4XX}+\frac{2}{3}\bar{G}_{5\phi X}+\frac{1}{2}(\bar{G}_{3X}-3\bar{G}_{4\phi X})^2-2(\bar{G}_{4X}-\bar{G}_{5\phi})^2\geq 0
\,, \label{Horn2}
\end{align}
which are also derived in~\cite{Melville:2019wyy}.\footnote{There are tiny mismatches between our results and the results presented in \cite{Melville:2019wyy}. For instance, the equations (1) and (2) in the supplemental material of \cite{Melville:2019wyy} contains the combination $\bar{G}_{4X}-\frac{1}{2}\bar{G}_{5\phi}$ while the combinations $\bar{G}_{4X}-\bar{G}_{5\phi}$ appears in our equations \eqref{Horn1} and \eqref{Horn2}. Since $\bar{G}_{4X}[XR+\{ (\Box \phi)^2-\phi_{\mu\nu}^2\} ]=\bar{G}_{4X}G^{\mu\nu}\partial_{\mu}\phi \partial_{\nu}\phi$ and $\bar{G}_{5\phi}\phi G^{\mu\nu}\nabla_{\mu}\nabla_{\nu}\phi=-\bar{G}_{5\phi}G^{\mu\nu}\partial_{\mu}\phi\partial_{\nu}\phi$ up to ignoring the boundary terms, the combination $\bar{G}_{4X}-\bar{G}_{5\phi}$ must appear. Except tiny mismatches of the coefficients, our results agree with the results of~\cite{Melville:2019wyy} (note that our definition of $G_3$ is opposite in sign to theirs and they use the normalization $\bar{f}_2=2$.).}

The discovery of the gravitational waves puts strong bounds on the ghost-free scalar-tensor theories. Let us suppose the DHOST theory is in the regime of validity even at the LIGO frequency scale (see the discussion of \cite{deRham:2018red}). The constraints from the speed of GWs~\cite{Langlois:2017dyl} and from the decay of GWs~\cite{Creminelli:2018xsv} then restrict the DHOST Lagrangian to
\begin{align}
	{\cal L} &= 
	    G_{2}(\phi,X) -G_{3}(\phi,X)\Box\phi
	   +f_2(\phi,X)R +\frac{3f_{2X}^2}{2f_2}(\partial_{\mu}X)^2
	   \,. \label{viableDHOST}
\end{align}
The Lagrangian \eqref{viableDHOST} is the viable model of the DHOST theory for dark energy. In addition, the cubic Galileon term $G_3\Box \phi$ induces ghost and/or gradient instabilities in a short scale in the presence of gravitational waves~\cite{Creminelli:2019kjy} which leads to a mild bound on the cubic Galileon term. Since this bound is not so stringent, we shall keep the $G_3$ term. The positivity bounds on \eqref{viableDHOST} is 
\begin{align}
&\frac{1}{2}\bar{G}_{2XX}-\bar G_{3X}(\bar G_{2\phi X} -2\bar G_{3\phi\phi})-\frac{1}{2}\bar{G}_{2\phi\phi}\bar G_{3X}^{2}
-2\bar{f}_{2X}(6\bar{f}_{2\phi\phi}+1)\geq 0
\,, \label{positivityDHOST}
\end{align}
where the second bound \eqref{DHOST2} is trivially satisfied in the viable model of DHOST. When the shift symmetry is assumed, the bound is reduced to $\bar{G}_{2XX}-4\bar{f}_{2X}\geq 0$.

The viable model should have a screening mechanism to pass the local tests of gravity. In the shift-symmetric theories, the screening can be realized by the Galileon terms $X\Box \phi$ and/or by non-linear kinetic terms like $X^2$. The former one is known as the Vainshtein screening~\cite{Vainshtein:1972sx}, and the latter one is often dubbed
as the kinetic screening~\cite{Babichev:2009ee}. There exists a viable parameter space to implement the Vainshtein screening and to satisfy the positivity bounds \cite{deRham:2017imi}. On the other hand, the positivity bounds may obstruct the kinetic screening because it requires a negative coefficient of the $X^2$ term~\cite{Brax:2012jr}.
The bound \eqref{positivityDHOST} prefers a positive sign of $X^2$ though $\bar{G}_{2XX}<0$ can be allowed due to the non-minimal coupling $\bar f_{2X}$ even in the shift-symmetric case. See \cite{Hirano:2019scf,Crisostomi:2019yfo} for discussions on the screening in the model \eqref{viableDHOST}.

\section{Conclusion and discussions}\label{sec:concl}
In this paper, we have studied the validity of the positivity bounds in the presence of a massless graviton, by considering the $2$ to $2$ scattering of a light scalar field $\phi$. We assumed several desired properties, such as analyticity and the $s^2$ boundedness \eqref{compfrois}, on the scattering amplitude. These properties are certainly satisfied in some known examples. We clarified that the imaginary part of the scattering amplitude grows as fast as $s^2$ in the forward limit.  
This is the essential and unavoidable origin of the possible violation of strict positivity in the presence of a massless graviton, even when the scattering amplitude is bounded at high energies as \eqref{compfrois}.
 
Assuming the Regge behavior of the amplitude, we explicitly computed the cancellation between the graviton $t$-channel pole and the UV integral of $\im\,F(s,-0)$, and obtained the finite residuals: the second and the third term in \eqref{subtdisp8}.
We then found that the Regge behavior and unitarity will be {\it insufficient} to derive the positivity bounds. 
We showed that the residuals can give negative corrections to the positivity bounds. Indeed they are {\it negative} in a well-known example, leading to the violation of strict positivity. We then obtained the upper bound on the allowed amount of the positivity violation. 

The upper bound is found to be suppressed by $\Mpl^{-2}\alpha'$ where $\alpha'$ is the scale of Reggeization which is typically determined by the string scale $M_{\rm s}$ as $\alpha'=M_{\rm s}^{-2}$ in string theory examples. This implies the utility of the positivity bounds when the cutoff scale of EFT is much lower than the scale of Reggeization, as is already argued in \cite{Hamada:2018dde}. We then applied the approximate positivity to scalar-tensor EFT at one-loop level. 

Implications on the DHOST theory, which is known as the most general ghost-free scalar tensor theory, were also discussed. The bounds are shown in \eqref{DHOST1} and \eqref{DHOST2}. Since this model has been widely discussed as phenomenological models in cosmology, our bounds could be useful to test the properties of UV completion of gravity from cosmological observations. 

If future observations verify the violation of our bounds on a given gravitational EFT, one will have to choose one of the following three options:
\begin{itemize}
\item [(1)] Gravitational theory at low energies is not described by such an EFT;
\item [(2)] UV completion of gravity does not exhibit the Regge behavior or violates the ``standard'' assumptions;
\item [(3)] UV completion of gravity exhibits the Regge behavior, but the Reggeization scale is sufficiently low so that the observed negativity can be compatible with our bound on the negativity \eqref{eq:negabd}.
\end{itemize}

There are several future directions from this work. Firstly, we expect that the bound on negativity \eqref{eq:negabd} might be improved by requiring more precise knowledge of UV completion. We may compute the first term of the r.h.s.~of \eqref{subtdisp8} depending on the models of the UV completions of gravity. If different UV completions give a different sign of the r.h.s.~of \eqref{subtdisp8}, observational constraints on the sign can be used to test the UV completion of gravity. Secondly, 
we should keep in mind that the positivity bounds are the constraints on the operators around the flat background, not around the cosmological background. It will be also necessary to carefully consider whether or not one can extend the positivity bounds on models around curved backgrounds in order to compare the bounds with the observations. We leave these aspects for future work.

\acknowledgments
We would like to thank Toshifumi Noumi for useful discussions. 
We would be especially grateful to Andrew J. Tolley for useful suggestions and valuable comments on this work.  J.\,T. was supported by the Japan Society for the Promotion of Science (JSPS) Postdoctoral Fellowship No. 202000912.
The work of K.A. was supported in part by Grants-in-Aid from the Scientific Research Fund of the Japan Society for the Promotion of Science, No.~19J00895 and No.~20K14468. S.H. is supported by JSPS KAKENHI Grant Numbers JP19H01895.


\appendix

\section{Extension to the positivity with $t$-derivative}\label{tderivext}
Firstly we briefly summarize the positivity bounds with $t$-derivatives for scalar theories without gravity in app.~\ref{tderivext1}, following \cite{deRham:2017avq}. We then discuss the approximate positivity for them in app.~\ref{tderivext2}.
\subsection{For scalar theories without gravity}\label{tderivext1}

By taking the derivative of \eqref{subtdisp3} with respect to $t$, one obtains
\begin{align}
&B^{(2N,M)}(t)\coloneqq\left.\frac{\der^{2N}_v\der^M_t}{M!}\tilde B(v,t)\right|_{v=0}=\frac{(2N)!2}{M!\pi}\int^\infty_{4m_\phi^2}\mathrm{d}\mu\,\der^M_t\left[\frac{\im\, F\left(\mu+i\epsilon,t\right)}{\left(\mu-2m_\phi^2+\frac{t}{2}\right)^{2N+1}}\right]
\end{align}
for $N=1,2,\cdots$ and $M=0,1,\cdots$ when $t<4m_\phi^2$. We can rewrite this expression as
\begin{align}
B_{\lth}^{(2N,M)}(t)
&=\sum_{k=0}^{M}\frac{(-1)^k}{k!2^k}I_{\lth}^{(2N+k,M-k)}\left(t\right)\,,\label{atchanposi1}
\end{align} 
where $B_{\lth}^{(2N,M)}(t)$ and $I^{(q,p)}_{\lth}(t)$ are defined by 
\begin{subequations}
\begin{align}
&B_{\lth}^{(2N,M)}(t)\coloneqq B^{(2N,M)}(t)-\frac{(2N)!2}{M!\pi}\int^{\lth^2}_{4m_\phi^2}\mathrm{d}\mu\,\der^M_t\left[\frac{\im\, F\left(\mu+i\epsilon,t\right)}{\left(\mu-2m_\phi^2+\frac{t}{2}\right)^{2N+1}}\right]\,,\\
&I_{\lth}^{(q,p)}\left(t\right)\coloneqq\frac{q!2}{p!\pi}\int^{\infty}_{\lth^2}\mathrm{d}\mu\left[\frac{\der_t^p\im\,F\left(\mu,t\right)}{\left(\mu-2m_\phi^2+\frac{t}{2}\right)^{q+1}}\right]>0\,.\label{aIdef1}
\end{align}
\end{subequations} 
It is known that eqs.~\eqref{atchanposi1} and \eqref{aIdef1} together with $\bigl(\lth^2+(t/2)-2m_\phi^2\,\bigr)I^{(q,p)}_{\lth}(t)<qI^{(q-1,p)}_{\lth}(t)$ lead to the following inequalities which are called improved positivity bounds for $N\geq1$, $M\geq0$, and $0\leq t<4m_\phi^2$ \cite{deRham:2017avq, deRham:2017imi}:
\begin{align}
Y_{\lth}^{(2N,M)}(t)&\coloneqq \sum_{r=0}^{M/2}c_rB_{\lth}^{(2N+r,M-2r)}(t)\no\\
&\quad+\frac{1}{\lth^2+(t/2)-2m_\phi^2}\sum_{{\rm even}\,k}^{(M-1)/2}\left(2(N+k)+1\right)\beta_kY_{\lth}^{\left(2(N+k),M-2k-1\right)}(t)>0\,,\label{aimprposi}
\end{align}
with $Y_{\lth}^{(2N,0)}(t)\coloneqq B_{\lth}^{(2N,0)}(t)$. Note that above inequalities with $N\geq2$ can be extended to $0\leq t<4m_\phi^2$. Coefficients $c_k$ and $\beta_k$ are recursively defined as
\begin{align}
&c_0=1\,,\quad c_k=-\sum_{r=0}^{k-1}\frac{2^{2(r-k)}c_r}{(2k-2r)!}\,\,\,\,(\forall k\geq1)\,,\quad \beta_k=(-1)^k\sum_{r=0}^k\frac{2^{2(r-k)-1}c_r}{(2k-2r+1)!}\,\,\,\,(\forall k\geq0)\,.\label{acoeff}
\end{align}
For example, eq.~\eqref{aimprposi} with $(2N,M)=(2,1)$ reads
\begin{align}
Y^{(2,1)}_{\lth}(t)=B^{(2,0)}_{\lth}(t)+\frac{3}{2\lth^2+t-4m_\phi^2}B^{(2,1)}_{\lth}(t)\simeq B^{(2,0)}_{\lth}(t)+\frac{3}{2\lth^2}B^{(2,1)}_{\lth}(t)>0\,,\label{abound21}
\end{align}
which is used in the main text.

\subsection{Cancellation of graviton $t$-channel pole and approximate positivity}\label{tderivext2}
Now we turn on the gravity. In sec.~\ref{correcttreat}, we have investigated an approximate positivity of $B_{\lth}^{(2,0)}(0)$. Generalizing the discussion in sec.~\ref{correcttreat}, we obtain the following twice-subtracted dispersion relation
\begin{align}
\frac{M!\pi}{4}B_{\lth}^{(2,M)}\left(0\right)
=&\int^{M^2_*}_{\lth^2}\mathrm{d}\mu\,\der_t^M\left[\frac{\im\, F\left(\mu,t\right)}{\left(\mu-2m_\phi^2+\frac{t}{2}\right)^{3}}\right]_{t=-0}\no\\
&+\der_t^M\left[D\left(t;M_*^2\right)+\der_v^{2}\left(\frac{\left.{\rm Res}_{t=0}F(s,t)\right|_{s=v+2m_\phi^2-(t/2)}}{-t-i\epsilon}\right)_{v=0}\right]_{t=-0}\,,\label{subtdisp4}
\end{align}
for $M\geq1$.
As is discussed in the main text, the terms in the second line in \eqref{subtdisp4} is suppressed by the Planck scale or the scale of Reggeization $\alpha'$. Typically this suppression will be of order of $\Mpl^{-2}{\alpha'}^{M+1}$. Therefore, when the cutoff scale of EFT is much lower than these scales so that the r.h.s.~of \eqref{subtdisp4} is dominated by the	first term, we have
\begin{subequations}
\begin{align}
&B_{\lth}^{(2,M)}\left(0\right)\approx\sum_{k=0}^{M}\frac{(-1)^k}{k!2^k}I_{\lth,M^2_*}^{(2+k,M-k)}\left(0\right)\,,\label{poledisp1}\\
&I_{\lth,M_*^2}^{(q,p)}\left(0\right)\coloneqq\frac{q!2}{p!\pi}\int^{M^2_*}_{\lth^2}\mathrm{d}\mu\lim_{t\to-0}\left[\frac{\der_t^p\im\,F\left(\mu,t\right)}{\left(\mu-2m_\phi^2+\frac{t}{2}\right)^{q+1}}\right]>0\,,
\end{align}
\end{subequations}
for $M\geq0$. We thus obtain the positivity bounds \eqref{aimprposi} with $N=1$ for $\forall M\geq0$, approximately. 

\section{Sub-leading correction: sample calculation}\label{reggesample}
In this appendix, we will evaluate how the sub-leading terms in the Reggeized amplitude of the form \eqref{regge2} contribute to $D\,\bigl(t;M_*^2\bigr)$. Firstly we consider polynomially suppressed sub-leading terms
\begin{align}
\im\, F(s,t)\simeq 
f(t)\left(\frac{\alpha's}{4}\right)^{2+j(t)}\left[1+\mathcal{O}\left(\left(\alpha's\right)^{-n}\right)\right]
\,,\label{regge5}
\end{align}
where $n$ denotes an arbitrary positive constant. These corrections give the tiny contributions to $D\,\bigl(t;M_*^2\bigr)$ compared to the leading term: for instance,
\begin{align}
\lim_{t\to-0}\int^\infty_{M_*^2}\frac{{\mathrm d}\mu}{\mu}\,(\alpha'\mu)^{j(t)-n}=\frac{1}{n\left(\alpha'M_*^2\right)^n}\ll1\,,
\end{align}
because $M_*^2\gg{\alpha'}^{-1}$. Next, we consider logarithmically suppressed sub-leading terms
\begin{align}
\im\, F(s,t)\simeq 
f(t)\left(\frac{\alpha's}{4}\right)^{2+j(t)}\left[1+\mathcal{O}\left(\ln^{-n}\left(\alpha's\right)\right)\right]
\,.\label{regge4}
\end{align}
Contributions from these sub-leading corrections to $D\,\bigl(t;M_*^2\bigr)$ can be computed as
\begin{align}
\int^\infty_{M_*^2}\frac{{\mathrm d}\mu}{\mu}\frac{(\alpha'\mu)^{j(t)}}{\ln^n(\alpha'\mu)}=\int^\infty_{\ln(M_*^2\alpha')}\mathrm{d}x\,\frac{e^{xj(t)}}{x^n}=\frac{\Gamma(-n+1,-j(t)\ln(M_*^2\alpha'))}{(-j(t))^{1-n}}\,,\label{eq:subleading1}
\end{align}
where $\Gamma(a,b)$ is the incomplete Gamma function. $0<n\leq 1$ case is inconsistent. For instance, when $n=1$,
\begin{align}
\int^\infty_{M_*^2}\frac{{\mathrm d}\mu}{\mu}\frac{(\alpha'\mu)^{j(t)}}{\ln(\alpha'\mu)}=\Gamma(0,-j(t)\ln(M_*^2\alpha'))\approx-\gamma-\ln(-j't)\,.
\end{align}
This contains $\ln(-t)$ divergence in the limit $t\to-0$, which is inconsistent with the validity of the twice-subtracted dispersion relation for $t<0$. For $0<n<1$, a singular term scaling as $t^{n-1}$ appears, which is again an inconsistent behavior. On the other hand, for $n>1$, eq.~\eqref{eq:subleading1} is finite in the limit $t\to-0$. Precise value can be evaluated as 
\begin{align}
\lim_{t\to-0}\int^\infty_{M_*^2}\frac{{\mathrm d}\mu}{\mu}\frac{(\alpha'\mu)^{j(t)}}{\ln^n(\alpha'\mu)}=\int^\infty_{\ln(M_*^2\alpha')}\mathrm{d}x\,\frac{1}{x^n}=\frac{1}{n-1}\ln^{n-1}(M_*^2\alpha')\,,
\end{align}
which gives a relevant contribution to $B^{(2,0)}_{\lth}(0)$, while there seems no reason to expect that the sign of these corrections could be fixed by unitarity etc. 

\section{One-loop corrections}
\label{sec:Xloop}
\begin{figure}[tbp]
 \centering
  \includegraphics[width=.3\textwidth, trim=250 270 360 120,clip]{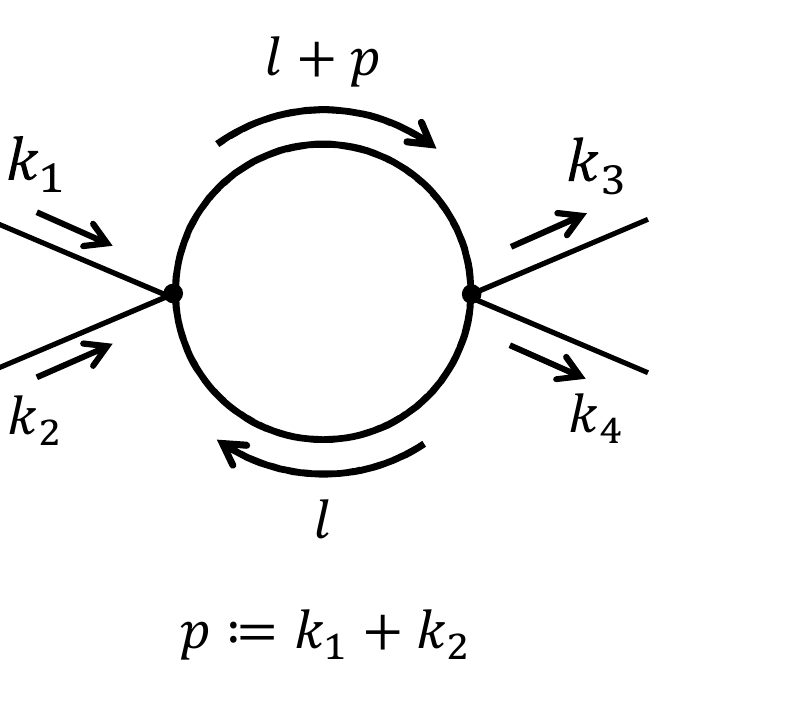}
 \caption{The $s$-channel loop diagram.}
 \label{loop} 
\end{figure}

Here we compute the loop corrections from the interaction vertex $\frac{{c_{44}'}}{4\Lambda^4}(\der\phi)^4$. The  $s$-channel diagram is shown in fig.~\ref{loop}. Its contribution, which is referred to as $I_{s-{\rm ch.}}(s,t)$, can be evaluated in $d$-dimension as 
\begin{align}
iI_{s-{\rm ch.}}(s,t)=&\frac{2(i{c_{44}'})^2}{\Lambda^8}\tilde\mu^{4-d}\int\frac{\mathrm{d}^dl}{(2\pi)^d}\tilde\Delta(l)\tilde\Delta(l+p)\left[-(k_1\cdot k_2)(l^2+l\cdot p)-k_1^\mu k_2^\nu\left(2l_\mu l_\nu+l_\mu p_\nu+l_\nu p_\mu\right)\right]\no\\
&\times\left[-(k_3\cdot k_4)(l^2+l\cdot p)-k_3^\rho k_4^\sigma\left(2l_\rho l_\sigma+l_\rho p_\sigma+l_\sigma p_\rho\right)\right]\,,\label{eq:s-loop1}
\end{align}
where $p\coloneqq k_1+k_2$ and $i\tilde\Delta(l)\coloneqq(l^2+m^2-i\epsilon)^{-1}$.  
We introduced a parameter $\tilde\mu$ with mass dimension 1 to keep ${c_{44}'}$ dimensionless. To make the expression \eqref{eq:s-loop1} simpler, we use the following useful equalities,
\begin{subequations}
\label{eq:formula}
\begin{align}
&\int\mathrm{d}^dk\,H\bigl(k^2\bigr)\,k^\mu=0\,,\quad\int\mathrm{d}^dk\,H\bigl(k^2\bigr)\,k^\mu k^\nu=\frac{g^{\mu\nu}}{d}\int\mathrm{d}k\,H\bigl(k^2\bigr)\,k^2\,,\\
&\int\mathrm{d}^dk\,H\bigl(k^2\bigr)\,k^\mu k^\nu k^\rho k^\sigma=\frac{1}{d(d+2)}\left(g^{\mu\nu}g^{\rho\sigma}+g^{\mu\rho}g^{\nu\sigma}+g^{\mu\sigma}g^{\nu\rho}\right)\int\mathrm{d}^dk\,H\bigl(k^2\bigr)\,\bigl(k^2\bigr)^2\,,
\end{align}
\end{subequations}
where $H\bigl(k^2\bigr)$ denotes an arbitrary Lorentz-scalar function of $k^2$. We can rewrite eq.~\eqref{eq:s-loop1} by using eqs.~\eqref{eq:formula} as
\begin{align}
iI_{s-{\rm ch.}}(s,t)=&\frac{2{c_{44}'}^2}{\Lambda^8}\tilde\mu^{4-d}\int\frac{\mathrm{d}^dl}{(2\pi)^d}i\tilde\Delta(l)i\tilde\Delta(l+p)\left[f(s,t)\,\bigl(l^2\bigr)^2+g(s)\,l^2\right]\,.\label{eq:s-loop2}
\end{align}
Here, $f(s,t)$ and $g(s)$ is defined by 
\begin{align}
&f(s,t)\coloneqq \left(\frac{1}{d}+\frac{1}{4}\right)(s-2m^2)^2+\frac{1}{d(d+2)}\left[(s-2m^2)^2+(t-2m^2)^2+(u(s,t)-2m^2)^2\right]\,,\\
&g(s)\coloneqq -\frac{s(s-m^2)^2}{d}\,.
\end{align}
By using the formula
\begin{align}
i\tilde\Delta(l)i\tilde\Delta(l+p)=\int^1_0\mathrm{d}x \frac{1}{\left(q^2+D(-p^2)-i\epsilon\right)^2}
\end{align}
with $q\coloneqq l+xp$ and $D(-p^2)\coloneqq (x^2-x)(-p^2)+m^2$,  we can perform the integration over $l$ in eq.~\eqref{eq:s-loop2} after the Wick rotation. By taking  $d=4-\varepsilon$ and expanding in terms of $\varepsilon$, we get
\begin{align}
I_{s-{\rm ch.}}(s,t)=&\frac{{c_{44}'}^2}{8\Lambda^8\pi^2}\left[\frac{1}{\varepsilon}I_0(s,t)+I_1(s,t)\right]+\mathcal{O}(\varepsilon)\,,\label{eq:s-loop4}
\end{align}
with
\begin{subequations}
\label{eq:s-loop5}
\begin{align}
I_0(s,t)&=(2m^2s+6m^4)f_0(s,t)-4m^2g_0(s)\,,\\
I_1(s,t)&=-\frac{2f_0(s,t)s^2}{15}+\left(2f_1(s,t)+\frac{2f_0(s,t)}{3}\right)m^2s+\frac{g_0(s)s}{6}-\left(g_0(s)-4g_1(s)\right)m^2+\frac{I_0}{2}\ln\mu^2\no\\
&-\int^1_0\mathrm{d}x\,\left[3f_0(s,t)D^2(s)+2\left(3f_0(s,t)sx^2-g_0(s)\right)D(s)+\left(f_0(s,t)s^2x^4-g_0(s)sx^2\right)\right]\ln D(s)\,.\label{eq:s-loop5b}
\end{align}
\end{subequations}
Here, functions $f_0(s,t)$, $f_1(s,t)$, $g_0(s)$, and $g_1(s)$ are defined by $f_0\coloneqq f|_{d=4}$, $f_1\coloneqq \lim_{\varepsilon\to0}(f|_{d=4-\varepsilon}-f_0)/\varepsilon$,  $g_0\coloneqq g|_{d=4}$, and $g_1\coloneqq\lim_{\varepsilon\to0}(g|_{d=4-\varepsilon}-g_0)/\varepsilon$. In eq.~\eqref{eq:s-loop5b}, we introduced a parameter $\mu$ which is defined by $\mu^2\coloneqq 4\pi e^{-\gamma}\tilde\mu^2$, and $\gamma$ is the Euler's constant. $\mu$ expresses a renormalization point. 
In the $\overline{\rm MS}$-scheme, contributions from counterterms exactly cancel the divergent term in eq.~\eqref{eq:s-loop4} that is proportional to $\varepsilon^{-1}$. Therefore, after the renormalization in this scheme, the one-loop correction from the vertex $\frac{{c_{44}'}}{4\Lambda^4}(\der\phi)^4$ is given by
\begin{align}
F_{\rm 1-loop}(s,t)=\frac{{c_{44}'}^2}{8\Lambda^8\pi^2}\left[I_1(s,t)+I_1(t,s)+I_1(u(s,t),t)\right]\,.\label{eq:Xloop1}
\end{align}
Here, the second and the third terms are the contributions from the $t$-channel diagram and the $u$-channel diagram, respectively. 

To consider improved positivity, we also need to evaluate the imaginary part. Taking the imaginary part of eq.~\eqref{eq:Xloop1} with imposing $s\geq 4m^2$, we get
\begin{align}
\im\,F_{\rm 1-loop}(s\geq 4m^2,t)
&=\frac{{c_{44}'}^2}{8\pi\Lambda^8}\sqrt{\frac{s-4m^2}{s}}\left[m^4f_0(s,t)-m^2g_0(s)\right]\,,\label{eq:loop-imag1}
\end{align}
which vanishes in the massless limit, interestingly.

\section{Loops, improved positivity, and the renormalization scale}\label{subsec:loop}

In sec.~\ref{application}, we discussed the sub-leading order positivity bound $B_{\lth}^{(4,0)}>0$ where the loop correction becomes important. However, there are several subtleties in general once loop corrections are taken into account. 
In this appendix, we will discuss how we can appropriately obtain  the bounds on the Wilson coefficients of EFT Lagrangian when loop corrections are taken into account. 

Once loop corrections are included, the Wilson coefficients of EFT Lagrangian run as one slides the renormalization scale $\mu$. 
Because of this renormalization group (RG) running, it is necessary to fix the renormalization scheme 
and the renormalization scale $\mu$ when expressing the positivity bounds in terms of the Wilson coefficients, in general. In this appendix, we shall use the dimensional regularization to regularize the UV divergences and adopt the $\overline{\rm MS}$-scheme, as we have done in the main text.  Then, what is an appropriate choice of the renormalization scale to formulate the positivity bounds? Physically speaking, an appropriate choice will be $\mu\sim \lth\gg m$, because non-renormalizable terms on which we would like to put non-trivial bounds will reflect the new-physics which become effective around/above the cutoff energy scale $\lth$. That is, it will be appropriate to express positivity bounds in terms of the scattering amplitudes evaluated at energy scales around $s\sim\lth^2$. We will confirm this in the following. 

To see what happens once loop corrections are included explicitly, let us consider the positivity bounds on the following classical effective Lagrangian which contains a renormalizable interaction
\begin{align}
\lag_{\rm cl}=-\frac{1}{2}(\der\phi)^2-\frac{m^2}{2}\phi^2- \frac{\lambda}{4!}\phi^4+\frac{c}{\Lambda^4}(\der\phi)^4\,,\label{eq:classicallag}
\end{align}
with $0<\lambda\ll1$, as a toy example. Here, we assume that the hierarchy between $m^2$ and  the cutoff scale $\Lambda^2$. This is just a toy model, but this model is good enough for discussing how one can take into account loop corrections when discussing the positivity bounds. At tree level, one has
\begin{align}
\left.\der_s^2F_{\rm tree}(s,0)\right|_{s=2m^2}=\frac{8c}{\Lambda^4}\,,\label{eq:tree1}
\end{align}
leading to the well-known bound $c>0$. Now let us consider the one-loop corrections to $F(s,0)$ up to $\mathcal{O}(\Lambda^{-4})$. The Lagrangian after adding the necessary counterterms is
\begin{align}
\lag=-\frac{1}{2}(\der\phi)^2-\frac{m^2}{2}\phi^2- \frac{Z_\lambda\lambda}{4!}\phi^4+\frac{Z_cc}{\Lambda^4}(\der\phi)^4\,.
\end{align}
Here, we neglect a field-renormalization and a mass-renormalization because they are irrelevant in our analysis below. It is important that the tree-level approximation \eqref{eq:tree1} is invalid in the presence of $\phi^4$ interaction: at the one-loop approximation one has
\begin{align}
\left.\der_s^2\left(F_{\rm tree}+F_{1-{\rm loop}}\right)(s,0)\right|_{s=2m^2}\simeq\frac{8c}{\Lambda^4}+\frac{\lambda^2}{32\pi^2m^4}\left(1-\frac{\pi}{4}\right)+\frac{\lambda c}{4\pi^2\Lambda^4}\left[-\ln\left(\frac{\mu^2}{m^2}\right)-2\pi+\frac{10}{3}\right]\,,\label{eq:1-loop1}
\end{align}
with choosing $Z_c=1+(16\pi^2\varepsilon)^{-1}\lambda$.\footnote{Counterterms $Z_\lambda$ is necessary to renormalize $F(s,0)$ at one-loop level, but its dependence vanishes after taking the derivative with respect to $s$. 
} Here, $\varepsilon$ denotes an infinitesimal parameter associated with the dimensional regularization procedure: we evaluated the l.h.s.~of \eqref{eq:1-loop1} in $d=4-\varepsilon$ dimensions and take the limit $\varepsilon\to0$ finally.
Eqs.~\eqref{eq:tree1} and \eqref{eq:1-loop1} show that 1-loop corrections from the renormalizable quartic coupling $\lambda\phi^4$ give the dominant contributions to  $\der_s^2 F(s,0)|_{s=2m^2}$, and hence the non-improved positivity bound $B^{(2,0)}(0)>0$ now reads
\begin{align}
B^{(2,0)}(0)\simeq\frac{8c}{\Lambda^4}+\frac{\lambda^2}{32\pi^2m^4}\left(1-\frac{\pi}{4}\right)+\frac{\lambda c}{4\pi^2\Lambda^4}\left[-\ln\left(\frac{\mu^2}{m^2}\right)-2\pi+\frac{10}{3}\right]>0\,,\label{eq:trivial}
\end{align}
which is trivially satisfied unless $\lambda$ is extremely suppressed to compensate the hierarchy between $m^2$ and $\Lambda^2$. The reason for this trivial positivity is that the effects from renormalizable interactions on $F(s,0)$ at low energies $s\sim m^2\ll\Lambda^2$ are much larger than those from non-renormalizable terms on which we would like to put non-trivial bounds. 
To improve this situation, let us consider the improved positivity bound
\begin{align}
B_{\lth}^{(2,0)}(0)
=&\frac{8c}{\Lambda^4}+\frac{\lambda^2}{32\pi^2m^4}\left(1-\frac{\pi}{4}\right)+\frac{\lambda c}{4\pi^2\Lambda^4}\left[-\ln\left(\frac{\mu^2}{m^2}\right)-2\pi+\frac{10}{3}\right]\no\\
&\quad-\frac{4}{\pi}\int^{\lth^2}_{4m^2}\mathrm{d}s'\,\frac{\im\,F\left(s'+i\epsilon,0\right)}{(s'-2m^2)^3}>0\,.\label{eq:impr1}
\end{align}
The imaginary part of $F(s,0)$, which is evaluated as 
\begin{align}
\im\,F(s+i\epsilon,0)&=
\frac{1}{32\pi}\left(\frac{s-4m^2}{s}\right)^{1/2}\left(\lambda^2-\frac{2\lambda c}{\Lambda^4}\left(s-2m^2\right)\left(s+8m^2\right)\right)\Theta(s-4m^2)+\mathcal{O}\left(\lambda^3\,,\,\frac{s^4}{\Lambda^8}\right)\,.\label{eq:imag1}
\end{align}
Here, $\Theta(z)$ denotes a Heaviside's step function. Using this expression, the term in the second line of eq.~\eqref{eq:impr1} can be evaluated as  
\begin{align}
\frac{4}{\pi}\int^{\lth^2}_{4m^2}\mathrm{d}s'\,\frac{\im\,F\left(s'+i\epsilon,0\right)}{(s'-2m^2)^3}\simeq\frac{\lambda^2}{32\pi^2m^4}\left(1-\frac{\pi}{4}\right)-\frac{\lambda^2}{16\pi^2\Lambda^4_{\rm th}}-\frac{\lambda c}{4\pi^2\Lambda^4}\left[\ln\left(\frac{\lth^2}{m^2}\right)+2\pi-5\right]\,,\label{eq:cut1}
\end{align}
where we neglected the terms suppressed by $(m^2/\lth^2)$ and the contributions from higher-order terms which are referred to as $\mathcal{O}\,\bigl(\lambda^3,\,(s^4/\Lambda^8)\bigr)$ in eq.~\eqref{eq:imag1}.   
Then, substituting eq.~\eqref{eq:cut1} into \eqref{eq:impr1}, the improved positivity bound $B_{\lth}^{(2,0)}(0)>0$ reads
\begin{align}
\frac{8c}{\Lambda^4}+\frac{\lambda^2}{16\pi^2\lth^4}-\frac{\lambda c}{4\pi^2\Lambda^4}\left[\ln\left(\frac{\mu^2}{\lth^2}\right)+\frac{5}{3}\right]>0\,.\label{eq:impr2}
\end{align} 
The third term is now tiny when choosing $\mu\approx\lth$. Specifically, we can choose $\mu=e^{-5/6}\lth$ to get\footnote{Note that the definition of $\lth$ implies that the modulus of the first term and the second term in \eqref{eq:improved1} will be the same order.}
\begin{align}
c(\mu)+\frac{\lambda^2}{16\pi^2}\left(\frac{\Lambda}{\lth}\right)^4+\mathcal{O}(\lambda^3)>0\,.\label{eq:improved1}
\end{align}
The situation here seems to be significantly improved compared to the one found in \eqref{eq:trivial}: the dominant contributions from renormalizable terms in \eqref{eq:trivial} are subtracted by considering the improved positivity bounds. Our result \eqref{eq:improved1} indicates that one can put meaningful constraints on the Wilson coefficients when taking $\mu\approx \lth$. 

\section{Inclusion of $R \phi$}
\label{sec:all_bound}
Let us put the operator,
\begin{align}
d_{10} \Mpl R \phi 
\,,
\end{align}
to the Lagrangian \eqref{lag2} with \eqref{scalarlag} which causes the mixing between the graviton and the scalar. One can perform field redefinitions to arrive at the Lagrangian \eqref{L_EFT}. Due to the mixing, $\mass$ is no longer the physical mass of $\phi$ when the fields are diagonalized. The physical mass in \eqref{L_EFT} is given by
\begin{align}
\mass'{}^2 = \frac{\mass^2}{1+6d_0^2}
\,.
\end{align}

The relevant coupling constants to the positivity bounds are $c_{44}',g_4'$ and $c_{48}'$  in \eqref{L_EFT}. 
In the presence of the $\phi R$ coupling, the relations are given by
\begin{align}
c_{44}'&=\frac{c_{44,1}}{(1+6d_{10}^2)^2} +\frac{c_{32}}{(1+6d_{10}^2)^3}\left( g_3 +6d_{10}(d_{22,1}+d_{22,2}) \frac{\Lambda^3}{\Mpl M^2} \right)
\no \\
&+\frac{\mass^2}{2(1+6d_{10}^2)^4\Lambda^2}\left( g_3 +6d_{10}(d_{22,1}+d_{22,2}) \frac{\Lambda^3}{\Mpl M^2} \right)^2
+\frac{2(1+12d_{20})}{(1+6d_{10}^2)^3} \left(d_{10}g_3 \frac{\Lambda}{\Mpl} -d_{22,2} \frac{\Lambda^4}{\Mpl^2M^2} \right)
\no \\
&+\frac{2d_{22,1} [1-d_{20}+16d_{10}^2(1+3d_{20})+24d_{10}^4]} {(1+6d_{10}^2)^3} \frac{\Lambda^4}{\Mpl^2M^2} 
\,.\\
g_4'&=\frac{g_4}{(1+6d_{10}^2)^2} +\frac{2d_{22,1}(d_{22,1}+d_{22,2})}{(1+6d_{10}^2)^2} \frac{\Lambda^6}{\Mpl^2 M^4}
-\frac{1}{2(1+6d_{10}^2)^3}\left( g_3 +6d_{10}(d_{22,1}+d_{22,2}) \frac{\Lambda^3}{\Mpl M^2} \right)^2
\,, \\
c_{48}'&=\frac{c_{48}}{(1+6d_{10}^2)^2}
\,.
\end{align}
One can then easily obtain the bounds in the presence of $\phi R$ from \eqref{leading_bound} and \eqref{subleading_bound}.

\bibliography{positivity_2020}

\end{document}